\documentclass[aps,prl,twocolumn,superscriptaddress]{revtex4-1}
\usepackage{graphicx}
\usepackage[colorlinks,bookmarks=false,citecolor=blue,linkcolor=blue,urlcolor=blue]{hyperref}
\usepackage{bm}
\usepackage{tikz}
\usepackage{amsmath,amssymb,amsfonts,amsthm}

\usepackage{pdfpages}
\makeatletter
\AtBeginDocument{\let\LS@rot\@undefined}
\makeatother

\makeatletter

\begin{document}
	
\title{Brillouin Platycosms and Topological Phases}
	
\author{Chen Zhang}
\affiliation{Department of Physics and HK Institute of Quantum Science \& Technology, The University of Hong Kong, Pokfulam Road, Hong Kong, China}

\author{Peiyuan Wang}
\affiliation{Department of Physics and HK Institute of Quantum Science \& Technology, The University of Hong Kong, Pokfulam Road, Hong Kong, China}

\author{Junkun Lyu}
\affiliation{Department of Physics and HK Institute of Quantum Science \& Technology, The University of Hong Kong, Pokfulam Road, Hong Kong, China}
	
\author{Y. X. Zhao}
\email[]{yuxinphy@hku.hk}
\affiliation{Department of Physics and HK Institute of Quantum Science \& Technology, The University of Hong Kong, Pokfulam Road, Hong Kong, China}

\begin{abstract}
	There exist ten distinct closed flat $3$D manifolds, known as platycosms, which hold significance in mathematics and have been postulated as potential geometric models for our universe. In this work, we demonstrate their manifestation as universes of Bloch particles, namely as momentum-space units referred to as Brillouin platycosms, which are natural extensions of the Brillouin torus within a broader framework of projective crystallographic symmetries. Moreover, we provide exact K-theoretical classifications of topological insulators over these platycosms by the Atiyah-Hirzebruch spectral sequence, and formulate a complete set of topological invariants for their identification. Topological phase transitions are generically characterized by Weyl semimetals, adhering to the generalized Nielsen-Ninomiya theorem: the total chirality number over a Brillouin platycosm is even (zero) if the platycosm is non-orientable (orientable). Our work generalizes the notion of Brillouin torus to ten Brillouin platycosms and therefore fundamentally diversifies the stages on which Block wavefunctions can perform their topological dance.
\end{abstract}
	
	\maketitle

\textcolor{blue}{\textit{Introduction}} 
Since 1933, it has been known that there exist a total of ten closed flat $3$D manifolds, named as platycosms~\cite{conway2003describing}. The ten platycosms are widely studied in the mathematics literature~\cite{thurston1997three,szczepanski2012geometry}, and have been speculated as potential shapes of a finite universe~\cite{Silk_PRL,Weeks_Nature}, inspired by the flatness of our universe at large scales~\cite{lachieze1995cosmic,levin2002topology,perlmutter1999measurements} and the fluctuation behavior of the cosmic microwave background. While it is still unsettled whether the physical universe is a platycosm, the universe for Bloch particles definitely is. It is just the $3$D Brillouin torus, namely the zeroth platycosm, in the ordinary theory of crystal symmetries. 

While the $3$D torus is just a cubic box with periodic boundary conditions for all three directions, the other nine platycosms have more intricate topologies. Four of them are non-orientable, since translating along some direction leads to a mirror reflection. They all  feature nontrivial $1$D cycles with finite orders. Particularly, all elementary $1$D cycles in the didicosm are fourfold, namely, traveling along any cycle four times necessarily leads to a contractible cycle, which inspired the popular science fiction story, \textit{Didicosm}~\cite{didicosm_novel}. The fascinating topological structures of all platycosms are pictorially illustrated in the Supplemental Materials (SM)~\cite{SM}.

In this work, we show that all the ten platycosms can appear as momentum-space units in the broader framework of projective crystal symmetries, and therefore refer them to as Brillouin platycosms.


Recently, projective crystal symmetries have been emerging as a research focus~\cite{PhysRevLett.126.196402,xue2022projectively,Fan_Zhang_PRL,li2022acoustic,meng2022spinful,PhysRevLett.132.236401,PhysRevLett.132.213801,PhysRevLett.130.236601,xiao2024revealing}. A particular feature is the existence of momentum-space nonsymmorphic symmetries~\cite{chen2022brillouin,PhysRevLett.130.256601}, which have been widely investigated in various artificial crystals~\cite{PhysRevLett.132.213801,PhysRevLett.132.266601,PhysRevB.109.134107,zhu2024brillouin,PhysRevApplied.21.044002,PhysRevA.109.013516,PhysRevB.108.L220101} and also naturally emerge in Moire systems~\cite{cualuguaru2025moire}, spin-space groups~\cite{xiao2024spin} and superconductors~\cite{PhysRevB.108.235412}.
Essentially different from symmorphic symmetries, a nonsymmorphic symmetry is a free transformation that shifts all points. Free transformations can further reduce the Brillouin torus into other platycosms. The reduction of the $2$D Brillouin torus to the Klein bottle by a momentum-space glide reflection resulting from a projective symmetry algebra has been demonstrated in Ref.~\cite{chen2022brillouin}. 

It is known that the ten platycosms $\mathcal{M}^\alpha$ have a one-to-one correspondence with the ten Bieberbach groups $B^\alpha$ with $\alpha=0,1,\cdots, 9$. They are the quotient manifolds of $\mathbb{R}^3$ under the actions of $B^\alpha$, i.e.,
\begin{equation}
	\mathcal{M}^\alpha = \mathbb{R}^3/B^\alpha.
\end{equation}
Thus, to realize the Brillouin platycosm $\mathcal{M}^\alpha$, we only need to realize the corresponding Bieberbach group $B^\alpha$ in momentum space $\mathbb{R}^3$. Every Bieberhach group $B^\alpha$ can be realized by an appropriate projective representation of a real-space symmorphic group $S^\alpha$~\cite{chen2023classification}. We have constructed physical tight-binding models and lattice Dirac models for all of them~\cite{SM}, paving the way for experimental realizations of the ten Brillouin platycosms.

\begin{table*}[!ht]
	\centering
	\includegraphics[width=\textwidth]{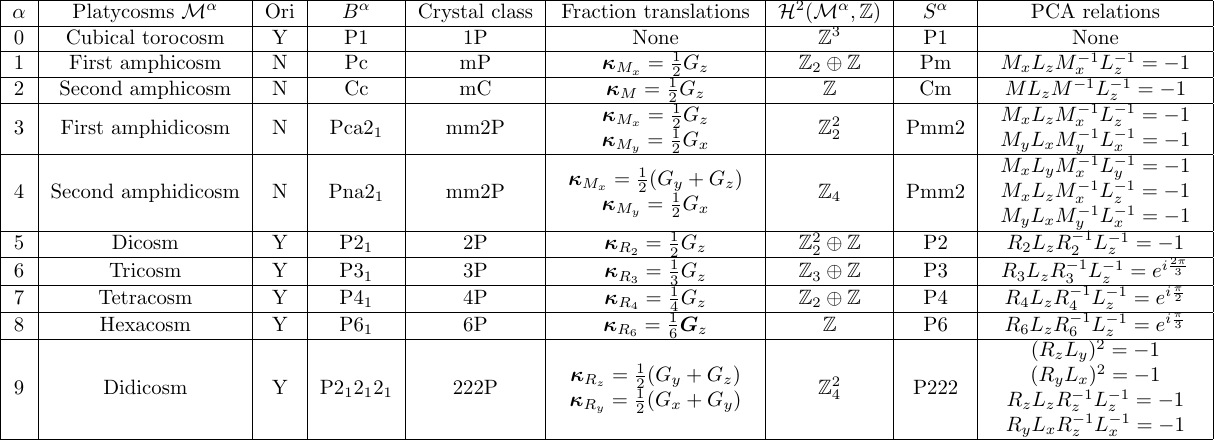}
	\caption{The first two columns list all ten platycosms $\mathcal{M}^{\alpha}$ with $\alpha = 0, \dots, 9$.  The orientability of $\mathcal{M}^\alpha$ is shown in the third column, where `$Y$' and `$N$' refer to orientable and non-orientable, respectively. The fourth column lists the corresponding Bieberbach groups $B^{\alpha}$. The following two columns exhibit their crystal classes and momentum space fractional translations, respectively. The reduced K-groups are listed in the seventh column. The corresponding real-space symmorphic groups $S^{\alpha}$ and projective symmetry algebraic relations are presented in the eighth and nine columns, respectively. 
}\label{tab:platycosms}
\end{table*}




Notably, the ten Brillouin platycosms exhaust all possible momentum units, as each crystallographic group has a maximal Berbiebach subgroup. Therefore, they are all the universes where Bloch wavefunctions can form abundant topological configurations.  By applying the Atiyah-Hirzebruch spectral sequence and the classifying space theory of groups, we show that for each platycosm the topological classification, namely the reduced K group, is isomorphic to the second cohomology group of the corresponding Bieberbach group, 
\begin{equation}\label{eq:K-H}
	\widetilde{K}(\mathcal{M}^\alpha)\cong H^2(B^\alpha,\mathbb{Z}).
\end{equation}
Based on this, we formulate a complete set of topological invariants for each classification. In general, transitions between two topological phases are characterized by a Weyl semimetal phase. These Weyl semimetals satisfy a generalized Nielsen-Ninomiya (NN) theorem~\cite{NIELSEN1981219}. Especially, the total chirality number over each non-orientable Brillouin platycosm can be any even integer.  


\textcolor{blue}{\textit{Realization of Brillouin platycosms}} 
To realize each Brillouin platyacosm $\mathcal{M}^\alpha$ as a momentum-space unit, we just need to realize the corresponding Bieberbach group $B^\alpha$ in momentum space~\cite{PhysRevLett.130.256601}. This is done by identifying the pair $(S^\alpha,\nu^\alpha)$ for each momentum-space $B^\alpha$. Here, $S^\alpha$ is the corresponding real-space symmorphic group and $\nu^\alpha$ the associated multiplier. 

Each Bieberbach group $B^\alpha$ belongs to an arithmetic crystal class $c^\alpha_F$, which specifies the lattice $L^\alpha_F$ or the primitive lattice translations for the translation subgroup, the point group $P^\alpha$, and how the point group acts on the lattice $L^\alpha_F$. 
Accordingly, $L^\alpha_F$ is the reciprocal lattice or the collection of all reciprocal translations. The corresponding real-space crystallographic group $S^\alpha$ is just the unique symmorphic group in the arithmetic crystal class $c^\alpha$ dual to $c^\alpha_F$, with the lattice $L^\alpha$ of $S^\alpha$ dual to $L^\alpha_F$~\cite{Dual_lattice}. 
For all $B^\alpha$, we tabulate their corresponding real-space symmorphic groups $S^\alpha$ in Tab. \ref{tab:platycosms}.

\begin{figure*}[!ht]
	\centering
	\includegraphics[width=\textwidth]{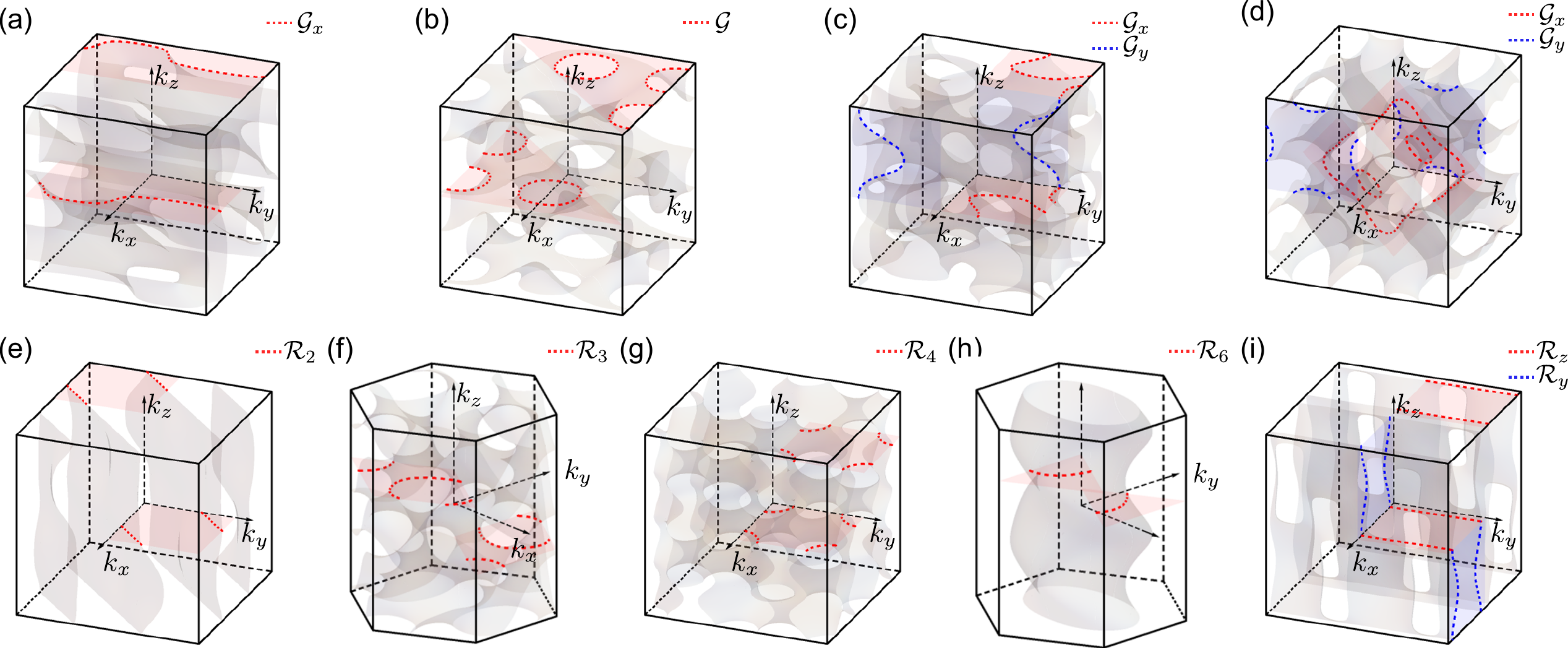}
	\caption{ Illustration of Brillouin platycosms via constant-energy surfaces. (a) to (i) correspond to $\alpha=1$ to $9$, respectively. In each sub-figure, the gray surfaces represent the constant-energy surfaces. Each blue or red dashed line indicates the intersection of these surfaces with a representative 2D plane, shaded with the same color. A pair of dashed lines of the same color on different 2D planes are related by a nonsymmorphic symmetry, as indicated in the upper-right corner.}\label{fig:energies}
\end{figure*}

In a projective representation $\rho^{\alpha}$ of $S^{\alpha}$ with multiplier $\nu^{\alpha}$, the group multiplication is modified to
$
\rho(g_1)\rho(g_2)=\nu^\alpha(g_1,g_2)\rho(g_1g_2).
$
According to Ref. \cite{PhysRevLett.130.256601}, the multiplier $\nu^\alpha$ for the momentum-space $B^\alpha$ is determined by the fractional reciprocal lattice translations $\bm{\kappa}_R$ through
\begin{equation}\label{eq:multiplier}
	\nu^\alpha[(\bm{t}_1,R_1),(\bm{t}_2,R_2)]=e^{-2\pi i\bm{\kappa}_{R_1}\cdot \bm{t}_2}.
\end{equation}
Here, each $g\in S^\alpha$ is denoted as $(\bm{t}, R)$ with $\bm{t}\in L^\alpha$ and $R$ in the point group $ P^\alpha$. 

It is significant to notice that six among the nine nontrivial platycosms, namely $\alpha=1,\cdots,5$ and $\alpha=9$, involve only half primitive translations and therefore the multipliers are valued in $\mathbb{Z}_2=\{\pm 1\}$. Consequently, $\mathbb{Z}_2$ gauge fluxes are sufficient to realize all the corresponding projective symmetry algebras. As has been demonstrated by numerous experiments, $\mathbb{Z}_2$ gauge fluxes can be flexibly engineered by artificial crystals~\cite{lu2014topological,huber2016topological,yang2015topological,ozawa2019topological, mittal2019photonic,Xue_2020, xue2021observation,ma2019topological,zhang2018topological, cooper2019topological, zhu2023topological,PhysRevA.110.023308,dalibard2011colloquium,prodan2009topological,imhof2018topolectrical, yu20204d, wang2023realization}. See the SM for a brief introduction~\cite{SM}. The remaining three Bieberbach groups, $\mathrm{Pn}_1$, requires $\mathbb{Z}_n$ gauge fields, with $n=3,4,6$.

In the SM~\cite{SM}, tight-binding models with a single state at each site and Dirac lattice models have been constructed for all momentum space $B^\alpha$ with appropriate gauge flux configurations, paving the way for experimental realizations. In Fig. \ref{fig:energies}, using the Dirac lattice models, we plot a constant energy surface within the Brillouin zone for each $B^\alpha$ to illustrate the topology of the corresponding platycosm.

\textcolor{blue}{\textit{Topological classification}}
We then proceed to the classification of topological insulators over the ten platycosms. For each $\mathcal{M}^\alpha$, the classification is given by the reduced K group $ \widetilde{K}(\mathcal{M}^\alpha)$. Each K group $K(\mathcal{M}^\alpha)$ can be analyzed by the Atiyah-Hirzebruch spectral sequence, which is stabilized at the fourth page. However, the spectral sequence cannot completely determine the K group.

In addition, each vector bundle gives rise to a determinant line bundle. As the complex line bundles are classified by the second cohomology group $\mathcal{H}^2(\mathcal{M}^\alpha)$~\cite{serre1955faisceaux,grothendieck1957quelques}, we have the natural group homomorphism,
\begin{equation}\label{eq:Det}
	\mathrm{Det}: \widetilde{K}(\mathcal{M}^\alpha)\rightarrow \mathcal{H}^2(\mathcal{M}^\alpha,\mathbb{Z}),
\end{equation}
which is apparently surjective. Since $\mathcal{M}^\alpha$ is the quotient space of $\mathbb{R}^3$ under the free action of $B^\alpha$, $\mathcal{M}^\alpha$ is the classifying space of $B^\alpha$.  Thus, $\mathcal{H}^2(\mathcal{M}^\alpha,\mathbb{Z})$ is isomorphic to the group-cohomology group $H^2(B^\alpha,\mathbb{Z})$,
\begin{equation}
	\mathcal{H}^2(\mathcal{M}^\alpha,\mathbb{Z})\cong H^2(B^\alpha,\mathbb{Z}).
\end{equation}
The advantage of this isomorphism is that the group-cohomology groups of $3$D space groups can be readily obtained from GAP ~\cite{GAP4}.

With the aid of the surjective homomorphism \eqref{eq:Det}, we can determine all K groups from the spectral sequences. The case-by-case study shows that $	\widetilde{K}(\mathcal{M}^\alpha)\cong H^2(B^\alpha,\mathbb{Z})$, i.e., Eq.~\eqref{eq:K-H}. All $\widetilde{K}(\mathcal{M}^\alpha)$ are listed in Tab.~\ref{tab:platycosms}, and all technical details for the above analysis can be found in the SM~\cite{SM}. 

\textcolor{blue}{\textit{Topological invariants}} It is significant to note that the isomorphism \eqref{eq:Det} in fact implies that all topological insulators can be characterized by the topologies of the corresponding determinant line bundles. In other words, all topological invariants can be formulated in terms of the Abelian Berry connection $a_\mu=\mathrm{Tr}\mathcal{A}_\mu$, with $\mathcal{A}_\mu$ the Berry connection of the valence bands.

From the reduced K groups, the topological invariant for each platycosm may be denoted as a vector $\nu=(\nu^{(n_1)}, \nu^{(n_2)}, \nu^{(n_3)})$. Here, $\nu^{(n)}\in Z_{n}$ with $\mathbb{Z}_{\infty}=\mathbb{Z}$ and $\mathbb{Z}_1=0$.

Each $\mathbb{Z}$-components just corresponds to the Chern numbers over some $2$D sub-tori. It is quite easy to identify which sub-tori can host nontrivial Chern numbers. For the first amphicosm $\mathcal{M}^1$ the Chern number is defined over the $k_y$-$k_z$ sub-torus as only on this sub-torus no momentum coordinates are reversed. For $\mathcal{M}^2$, the sub-torus is spanned by the anti-diagonal line on the $k_x$-$k_y$ plane and the reciprocal lattice vector along the $k_z$ direction. For $\mathcal{M}^5,\mathcal{M}^6,\cdots,\mathcal{M}^8$, the torus for the Chern number is spanned by the two primitive reciprocal lattice vectors on the $k_x$-$k_y$ plane.

The $\mathbb{Z}_n$ components can be formulated by the method used by Dijkgraaf and Witten to formulate the Chern-Simons theories for finite groups~\cite{dijkgraaf1990topological}. For each $\mathbb{Z}_n$, we choose a $2$-chain $c$ so that a set of symmetries $B^\alpha$ freely acts on its boundary $\partial c$ and therefore naturally divide $\partial c$ into $n$ symmetry-related $1$-chains. Note that in general, $c$ is a $\mathbb{Z}$-linear combination of sub-manifolds in momentum space. Because of the free action, we can require that the Bloch wavefunctions $\psi_a$ at symmetry-related momenta are the same, i.e.,
\begin{equation}
	\psi_a(\bm{k})=\psi_a(g\bm{k}),
\end{equation}
where $g$ is an arbitrary symmetry and $\bm{k}$ any point in momentum space. Then, the Abelian Berry connection $a$ is invariant under the Bieberbach group. Under this gauge condition, the $\mathbb{Z}_n$ invariant is formulated as
\begin{equation}
	\nu^{(n)}=\frac{1}{2\pi} \int_{c}f-\frac{n}{2\pi} \int_{\partial c/B^\alpha} a \mod n.
\end{equation}
Here, $f$ denotes the Abelian Berry curvature, and $\partial c/B^\alpha$ may be represented by any $1$-chain divided from the action of $B^\alpha$. The second term is equal to the Berry phase over $\partial c$ divided by $2\pi$, and therefore $\nu$ is valued in integers. As a different choice of the phases of the Bloch wavefunctions may change the Berry phase by $2\pi$, only $\nu \mod n$ is gauge invariant.


\begin{figure}[!t]
	\includegraphics[width=\columnwidth]{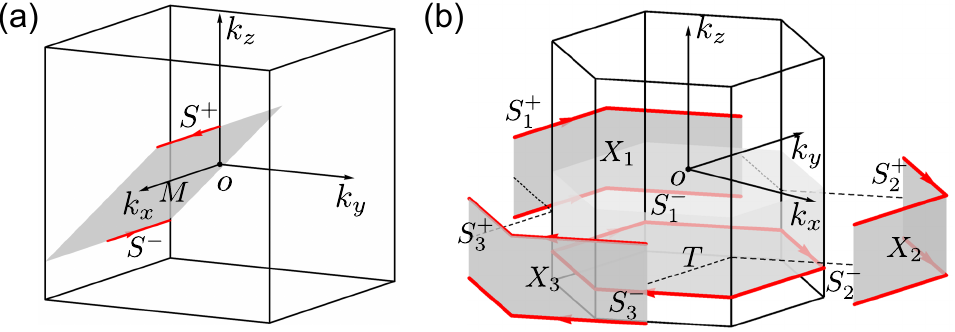}
	\caption{Illustration of the topological invariant formulation. (a) and (b) correspond to Bieberbach groups $Pna2_1$ and $P3_1$, respectively. The red lines indicate the oriented boundaries of the $2$D submanifolds marked in gray. In (b), the drawing of $X_i$ with $i=1,2,3$--representing the three pieces of the hexagonal prism's surface--is extended outward for clarity.}\label{fig:invariants}
\end{figure}

Following this general method, a complete set of topological invariants for all classifications are  formulated in the SM~\cite{SM}. As observed from Tab.~\ref{tab:platycosms}, the torsion components exhibit types $\mathbb{Z}_2$, $\mathbb{Z}_4$ and $\mathbb{Z}_3$. 

Each $\mathbb{Z}_2$ topological invariant just resembles that over the Klein bottle~\cite{chen2022brillouin}, i.e., one can always find a $2$D sub-manifold with the Klein-bottle topology under the action of the Bieberbach group. It is interesting to note that the $\mathbb{Z}_2$-component of the first amphicosm disappears in the topological invariant of the second amphicosm. In the later case, the $\mathbb{Z}_2$-invariant over the Klein bottle formed by the diagonal glide reflection equals the Chern number mod $2$~\cite{SM}. 
	
Two platycosms, second amphidicosm and didicosm, have $\mathbb{Z}_4$ components in their classifications. Here, we choose second amphidicosm as an example, since the classification is just $\mathbb{Z}_4$. The corresponding Bieberbach group is $Pna2_1$. Except for the reciprocal translation along the $k_z$ direction, $Pna2_1$ has two more generators, namely the nonsymmorphic symmetries,
\begin{equation}
	\begin{split}
	\mathcal{G}_x:~(k_x,k_y,k_z)&\rightarrow(-k_x,k_y+G_y/2,k_z+G_z/2),\\
	\mathcal{G}_y:~(k_x,k_y,k_z)&\rightarrow(k_x+G_x/2,-k_y,k_z).
	\end{split}
\end{equation}
Here, $G_{i}$ denotes the reciprocal lattice length along the $k_i$ direction with $i=x,y,z$. 

To formulate a $\mathbb{Z}_4$ topological invariant, we manage to choose the $2$D sub-manifold $M$ as plotted in Fig.~\ref{fig:invariants}(a).  $M$ is a sloped rectangle along the diagonal direction in the $k_y$-$k_z$ plane, spans the $k_x$ direction and a half of the diagonal direction in the $k_y$-$k_x$ plane, with the $k_x$ axis in the middle. It assumes the periodic boundary conditions along the $k_x$ direction as imposed by the reciprocal translation along the $k_x$ direction. The two glide reflection symmetries $\mathcal{G}_{x,y}$ act nontrivially on the boundary $\partial M$ of $M$, namely the two $k_x$-edges. The fundamental domain $\partial M/G$ is represented by the oriented segment $S^+$ in Fig.~\ref{fig:invariants}(a).  One can trace the orbit of $S^+$ under the transformation of $\mathcal{G}_x$ and $\mathcal{G}_y$ to verify that $S^+$ does generate the boundary $\partial M$. Thus, the topological invariant is given by
\begin{equation}\label{eq:z4_inv}
	\nu^{(4)}=\frac{1}{2\pi}\int_{X}f-\frac{2}{\pi}\int_{S^+} a \mod 4.
\end{equation}

The $\mathbb{Z}_3$ component appears in the topological classification for tricosm. The corresponding Bieberbach group is $\mathrm{P3}_1$, which is generated by reciprocal translations on the $k_x$-$k_y$ plane and the screw rotation $\mathcal{R}_3$.
$\mathcal{R}_3$ rotates the momentum-component $\bm{k}_{\perp}$ on the $k_x$-$k_y$ plane by $2\pi/3$ and translates $k_z$ by $G_z/3$, i.e.,
\begin{equation}
	\mathcal{R}_3:(\bm{k}_{\perp},k_z)\rightarrow (R_3\bm{k}_{\perp},k_z+G_z/3).
\end{equation}
To formulate a $\mathbb{Z}_3$ topological invariant, let us consider three cylinders $X_i$ with $i=1,2,3$ as illustrated in Fig.~\ref{fig:invariants}(b). The boundary $\partial X_i$ consists of two oriented components $S_i^{\pm}$ with $i=1,2,3$ and the screw rotation moves $S_i^+$ to $S^+_{i+1}$ with $i \mod 3$.
Accordingly, the $\mathbb{Z}_3$ topological invariant is given by
\begin{equation}
	\nu^{(3)}=\frac{1}{2\pi}\int_{X_1-X_2} f+\frac{3}{2\pi}\int_{S_1^{-}} a \mod 3.
\end{equation}
More technical details can be found in the SM~\cite{SM}.

In general, the formulation of a topological invariant is not unique. Relations between some typical formulations are discussed in the SM~\cite{SM}.

\textcolor{blue}{\textit{Criticality and the generalized Nielsen-Ninomiya theorem}} 
The critical phase between any two topological phases over a Brillouin platycosms is generically a Weyl semimetal. This can be attributed to the fact that no crystal symmetries exist at any point in a Brillouin platycosm, since the action of the Bieberbach group on momentum space is free. Consequently, crossing points of energy bands are generically twofold degenerate Weyl points. This has been thoroughly demonstrated by lattice models in the SM~\cite{SM}.

Recently, the NN theorem has been generalized from the Brillouin torus to the Brillouin first amphicosm, meaning that the total chirality number of all Weyl points over the first amphicosm can be any even integer and is not necessarily zero~\cite{PhysRevLett.132.266601}. In fact, the NN theorem can be readily extended to all Brillouin platycosms, with the result depending solely on their orientability. For all orientable Brillouin platycosms, the total chirality number of Weyl points equals zero, while for non-orientable ones, the total chirality number can be any even integer.

 The NN theorem can be rigorously demonstrated by applying the Poincaré-Hopf theorem to the Berry-curvature vector field $\bm{\Omega}=\nabla\times\bm{a}$~\cite{hopf2001vektorfelder}. The theorem asserts that, for each orientable platycosm, the sum of the indices (chirality numbers) of the zeros (Weyl points) equals the Euler characteristic, which is zero for all orientable platycosms~\cite{Vec_notes}. While all non-orientable platycosms also have a null Euler characteristic, the Poincaré-Hopf theorem holds only modulo 2 for non-orientable platycosms due to the absence of a global orientation~\cite{stiefel}. This can be understood as follows: on a non-orientable platycosm, starting from any Weyl point, there exists a closed cycle $\gamma$, and moving the Weyl point along $\gamma$ inverts its chirality number. As a result, for all non-orientable platycosms, the sum of chirality numbers of Weyl points equals zero modulo 2.

\textcolor{blue}{\textit{Summary and discussion}} In summary, we have shown that the ten Brillouin platycosms are momentum-space units for projective crystal symmetries, derived the K-theoretical classifications of topological insulators as the second cohomology groups of the corresponding Bieberbach groups, formulated all the topological invariants and investigated the critical Weyl semimetals.

In general, the boundary states of these multi-component topological invariants are difficult to specify in detail. For the cubical torocosm, namely the $3$D torus, the fascinating surface braidings of the Chern vectors were recently observed~\cite{Baile_Nature}.  Some attempts were made for the first amphicosm and the first amphidicosm~\cite{PhysRevLett.132.266601,PhysRevB.109.134107,zhu2024brillouin}, showing nontrivial twists of surface states. It is expected that with the complete topological classifications, more fascinating boundary-state structures can be found for topological phases over Brillouin platycosms.
\bibliographystyle{apsrev4-1}
\bibliography{references}

\begin{thebibliography}{65}
\bibitem [{Dua()}]{Dual_lattice}%
  \BibitemOpen
  \href@noop {} {}\bibinfo {note} {The primitive lattice vectors $\bm{e}_i$ and $\bm{G}_i$ for $L$ and $L_F$, respectively, are related by $\bm{e}_i=\frac{1}{2}\epsilon_{ijk}\bm{G}_j\times \bm{G}_k/|\bm{G}_1\cdot(\bm{G}_2\times\bm{G}_3)|$}\BibitemShut {NoStop}%
\bibitem [{Vec()}]{Vec_notes}%
  \BibitemOpen
  \href@noop {} {}\bibinfo {note} {Rigorously speaking, Weyl points are singular points of the vector field $\bm{\Omega}$. For each Weyl point, we can apply a radial rescaling in a neighborhood disk centered at the singular point, which leads to a smooth vector field. Then, all singular points become zeros, and the chirality number of each Weyl point equals the index of the zero.}\BibitemShut {Stop}%
\end{thebibliography}
\widetext
\clearpage


\section*{Supplementary material}

# Supplemental Materials for "Brillouin Platycosms and Topological Phases"

## Contents

## I. Homology and cohomology groups of platycosms

The classification of line bundles of valence bands over Brillouin platycosm $\mathcal{M}^\alpha$ is given by the cohomology group $\mathcal{H}^2(\mathcal{M}^\alpha, \mathbb{Z})$. Since $\mathcal{M}^\alpha$ is the classifying space of $B^\alpha$, $\mathcal{H}^n(\mathcal{M}^\alpha, \mathbb{Z})$ is isomorphic to the group-cohomology group $H^n(B^\alpha, \mathbb{Z})$,

$$\mathcal{H}^n(\mathcal{M}^\alpha, \mathbb{Z}) \cong H^n(B^\alpha, \mathbb{Z}). \tag{S1}$$

The isomorphism is also held for homology groups, i.e.,

$$\mathcal{H}_n(\mathcal{M}^\alpha, \mathbb{Z}) \cong H_n(B^\alpha, \mathbb{Z}). \tag{S2}$$

From the universal coefficient theorem of cohomology groups, we know

$$\mathcal{H}^2(\mathcal{M}^\alpha, \mathbb{Z}) \cong T_1^\alpha \oplus F_2^\alpha, \tag{S3}$$

here $T_1^\alpha$ is the torsion component of $H_1(B^\alpha, \mathbb{Z})$, and $F_2^\alpha$ is the free Abelian group component $\mathbb{Z}^{n_\alpha}$ of $H_2(B^\alpha, \mathbb{Z})$. The following table lists the first three homology groups and cohomology groups. Results are obtained from GAP.

| $\alpha$ | $B^\alpha$ | $\mathcal{H}_1(\mathcal{M}^\alpha)$ | $\mathcal{H}_2(\mathcal{M}^\alpha)$ | $\mathcal{H}_3(\mathcal{M}^\alpha)$ | $\mathcal{H}^1(\mathcal{M}^\alpha)$ | $\mathcal{H}^2(\mathcal{M}^\alpha)$ | $\mathcal{H}^3(\mathcal{M}^\alpha)$ | Presentation |
|---|---|---|---|---|---|---|---|---|
| 0 | P1 | $\mathbb{Z}^3$ | $\mathbb{Z}^3$ | $\mathbb{Z}$ | $\mathbb{Z}^3$ | $\mathbb{Z}^3$ | $\mathbb{Z}$ | $\langle L_x, L_y, L_z | [L_x, L_y], [L_x, L_z], [L_y, L_z] \rangle$ |
| 1 | Pc | $\mathbb{Z}_2 \oplus \mathbb{Z}^2$ | $\mathbb{Z}_2 \oplus \mathbb{Z}$ | 0 | $\mathbb{Z}^2$ | $\mathbb{Z}_2 \oplus \mathbb{Z}$ | $\mathbb{Z}_2$ | $\langle L_x, L_y, \mathcal{G}_x | \mathcal{G}_x L_x \mathcal{G}_x^{-1} = L_x^{-1}, [L_x, L_y], [\mathcal{G}_x, L_y] \rangle$ |
| 2 | Cc | $\mathbb{Z}^2$ | $\mathbb{Z}_2 \oplus \mathbb{Z}$ | 0 | $\mathbb{Z}^2$ | $\mathbb{Z}$ | $\mathbb{Z}_2$ | $\langle L_a, L_b, \mathcal{G} | \mathcal{G} L_a \mathcal{G}^{-1} = L_b, [L_a, L_b] \rangle$ |
| 3 | Pca$2_1$ | $\mathbb{Z}_2^2 \oplus \mathbb{Z}$ | $\mathbb{Z}_2$ | 0 | $\mathbb{Z}$ | $\mathbb{Z}_2^2$ | $\mathbb{Z}_2$ | $\langle L_y, \mathcal{G}_x, \mathcal{G}_y | \mathcal{G}_x L_y \mathcal{G}_x^{-1} = \mathcal{G}_y^{-1}, \mathcal{G}_y L_y \mathcal{G}_y^{-1} = L_y^{-1}, [\mathcal{G}_x, L_y] \rangle$ |
| 4 | Pna$2_1$ | $\mathbb{Z}_4 \oplus \mathbb{Z}$ | $\mathbb{Z}_2$ | 0 | $\mathbb{Z}$ | $\mathbb{Z}_4$ | $\mathbb{Z}_2$ | $\langle L_y, \mathcal{G}_x, \mathcal{G}_y | \mathcal{G}_x \mathcal{G}_y \mathcal{G}_x^{-1} = L_y \mathcal{G}_y^{-1}, \mathcal{G}_y L_y \mathcal{G}_y^{-1} = L_y^{-1}, [\mathcal{G}_x, L_y] \rangle$ |
| 5 | P$2_1$ | $\mathbb{Z}_2^2 \oplus \mathbb{Z}$ | $\mathbb{Z}$ | $\mathbb{Z}$ | $\mathbb{Z}$ | $\mathbb{Z}_2^2 \oplus \mathbb{Z}$ | $\mathbb{Z}$ | $\langle L_x, L_y, \mathcal{R}_2 | \mathcal{R}_2 L_x \mathcal{R}_2 = L_x^{-1}, \mathcal{R}_2 L_y \mathcal{R}_2 = L_y^{-1}, [L_x, L_y] \rangle$ |
| 6 | P$3_1$ | $\mathbb{Z}_3 \oplus \mathbb{Z}$ | $\mathbb{Z}$ | $\mathbb{Z}$ | $\mathbb{Z}$ | $\mathbb{Z}_3 \oplus \mathbb{Z}$ | $\mathbb{Z}$ | $\langle L_a, L_b, \mathcal{R}_3 | \mathcal{R}_3 L_a \mathcal{R}_3^{-1} = L_a, \mathcal{R}_3 L_b \mathcal{R}_3^{-1} = L_a^{-1} L_b^{-1}, [L_a, L_b] \rangle$ |
| 7 | P$4_1$ | $\mathbb{Z}_2 \oplus \mathbb{Z}$ | $\mathbb{Z}$ | $\mathbb{Z}$ | $\mathbb{Z}$ | $\mathbb{Z}_2 \oplus \mathbb{Z}$ | $\mathbb{Z}$ | $\langle L_x, L_y, \mathcal{R}_4 | \mathcal{R}_4 L_x \mathcal{R}_4^{-1} = L_y, \mathcal{R}_4 L_y \mathcal{R}_4^{-1} = L_x^{-1}, [L_x, L_y] \rangle$ |
| 8 | P$6_1$ | $\mathbb{Z}$ | $\mathbb{Z}$ | $\mathbb{Z}$ | $\mathbb{Z}$ | $\mathbb{Z}$ | $\mathbb{Z}$ | $\langle L_a, L_b, \mathcal{R}_6 | \mathcal{R}_6 L_a \mathcal{R}_6^{-1} = L_b, \mathcal{R}_6 L_b \mathcal{R}_6^{-1} = L_a^{-1} L_b, [L_a, L_b] \rangle$ |
| 9 | P$2_1 2_1 2_1$ | $\mathbb{Z}_4^2$ | 0 | $\mathbb{Z}$ | 0 | $\mathbb{Z}_4^2$ | $\mathbb{Z}$ | $\langle \mathcal{R}_y, \mathcal{R}_z | \mathcal{R}_y = \mathcal{R}_z^2 \mathcal{R}_y \mathcal{R}_z^2, \mathcal{R}_z = \mathcal{R}_y^2 \mathcal{R}_z \mathcal{R}_y^2 \rangle$ |

TABLE S1: he first two columns list the labels and the ten Bieberbach groups, respectively. The next three columns, labeled $\mathcal{H}_1(\mathcal{M}^\alpha)$, $\mathcal{H}_2(\mathcal{M}^\alpha)$, and $\mathcal{H}_3(\mathcal{M}^\alpha)$, present the first three homology groups of the ten platycosms. The following three columns display the cohomology groups $\mathcal{H}^1(\mathcal{M}^\alpha)$, $\mathcal{H}^2(\mathcal{M}^\alpha)$, and $\mathcal{H}^3(\mathcal{M}^\alpha)$. The final column provides the explicit presentations of the corresponding Bieberbach groups.

## II. The derivation of K groups by the Atiyah–Hirzebruch spectral sequence

The reduced K-groups $\widetilde{K}(\mathcal{M}^\alpha)$ provide the stable classification of topological insulators over the ten Brillouin platycosms. Each K-group $K(\mathcal{M}^\alpha)$ can be analyzed through the Atiyah-Hirzebruch spectral sequence. For each $\mathcal{M}^\alpha$, consider a spectral sequence $\{E_r^{p,q}(\mathcal{M}^\alpha), d_r\}_{r \geq 1}$ with the first page given by

$$E_1^{p,q}(\mathcal{M}^\alpha) \cong \mathcal{C}^p(\mathcal{M}^\alpha, K^q(pt)), \tag{S4}$$

where $\mathcal{C}^p(\mathcal{M}^\alpha, K^q(pt))$ denotes the $p$-cochains with coefficients in $K^q(pt)$, and the differential $d_1$ is the standard coboundary operator. The second page is given by

$$E_2^{p,q}(\mathcal{M}^\alpha) \cong \mathcal{H}^p(\mathcal{M}^\alpha, K^q(pt)), \tag{S5}$$

which represents the $p$-th cohomology of $\mathcal{M}^\alpha$ with coefficients in $K^q(pt)$. The K-groups of a single point $K^q(pt)$ satisfy

$$K^q(pt) \cong \begin{cases} \mathbb{Z}, & \text{if } q \text{ is even}, \\ 0, & \text{if } q \text{ is odd}. \end{cases} \tag{S6}$$

We adopt the notation $K(\mathcal{M}^\alpha) \cong K^0(\mathcal{M}^\alpha)$. The spectral sequence evolves according to the differential

$$E_{r+1}^{p,q} = \frac{\ker d_r : E_r^{p,q} \to E_r^{p+r,q-r+1}}{\operatorname{im} d_r : E_r^{p-r,q+r-1} \to E_r^{p,q}}, \tag{S7}$$

where $d_r : E_r^{p,q} \to E_r^{p+r,q-r+1}$ vanishes for even $r$, as $E_r^{p,q} \cong 0$ for all odd values of $q$. Therefore, there are no updates from $E_2$ to $E_3$, i.e., $E_2^{p,q} \cong E_3^{p,q}$ for all $p$ and $q$.

The $E_2$ and $E_3$ pages are depicted in Fig. S1, where the horizontal axis corresponds to $p$ and the vertical axis to $q$. Pages with $p \geq 4$ or $p \leq -1$ are omitted, as all elements in these pages are zero. Note that the $E_2$ and $E_3$ pages exhibit a periodicity in $q$ with period 2.

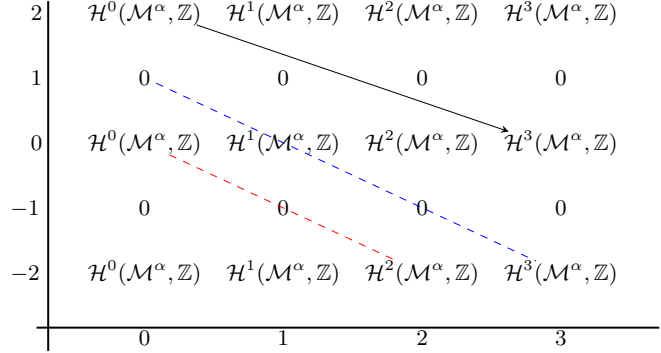

FIG. S1: $E_2$ and $E_3$ pages of $\mathcal{M}^\alpha, \alpha = 0, \ldots, 9$. The horizontal axis corresponds to $p$ and the vertical axis to $q$. The red dashed line represents filtration of $K^0(\mathcal{M}^\alpha)$ and the blue dashed line represents filtration of $K^1(\mathcal{M}^\alpha)$.

When updating from the $E_3$ to the $E_4$ page, it is sufficient to consider the following chain complex:

$$0 \to \mathcal{H}^0(\mathcal{M}^\alpha, \mathbb{Z}) \to \mathcal{H}^3(\mathcal{M}^\alpha, \mathbb{Z}) \to 0, \tag{S8}$$

which is indicated by the arrow in Fig. S1. We need to treat the cases of orientable and non-orientable platycosms separately. The zero cohomology group $\mathcal{H}^0(\mathcal{M}^\alpha, \mathbb{Z})$ is isomorphic to $\mathbb{Z}$ for all platycosms. The third cohomology group $\mathcal{H}^3(\mathcal{M}^\alpha, \mathbb{Z})$ is isomorphic to $\mathbb{Z}$ if $\mathcal{M}^\alpha$ is orientable, and is isomorphic to $\mathbb{Z}_2$ if $\mathcal{M}^\alpha$ is non-orientable. Since platycosms are 3-manifolds, the spectral sequence stabilizes at the fourth page.

For orientable $\mathcal{M}^\alpha$ with $\alpha = 0, 5, \ldots, 9$, the chain complex in Eq. (S8) takes the form:

$$0 \to \mathbb{Z} \xrightarrow{d} \mathbb{Z} \to 0. \tag{S9}$$

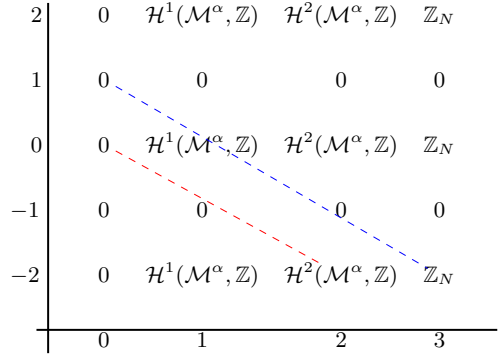

FIG. S2: $E_4$ pages of orientable $\mathcal{M}^\alpha$ with nontrivial map $d$ and $\alpha = 0, 5, \ldots, 9$. The horizontal axis corresponds to $p$ and the vertical axis to $q$. The red dashed line represents filtration of $K^0(\mathcal{M}^\alpha)$ and the blue dashed line represents filtration of $K^1(\mathcal{M}^\alpha)$.

In this case, the map $d$ sends $1 \in \mathbb{Z}$ to some integer $N \in \mathbb{Z}$. When $N = 0$, no update occurs on the $E_4$ page, i.e., $E_4^{p,q} \cong E_3^{p,q}$. However, if $N \neq 0$, the $E_4$ page is updated as shown in Fig. S2. However, this case can be ruled out. By filtration, we obtain the group extension of $K^0(\mathcal{M}^\alpha)$:

$$0 \to 0 \to K^0(\mathcal{M}^\alpha) \to \mathcal{H}^2(\mathcal{M}^\alpha, \mathbb{Z}) \to 0. \tag{S10}$$

This implies that $\widetilde{K}(\mathcal{M}^\alpha) \oplus \mathbb{Z} \cong K^0(\mathcal{M}^\alpha) \cong \mathcal{H}^2(\mathcal{M}^\alpha, \mathbb{Z})$, which contradicts the fact that the map $\text{Det} : \widetilde{K}(\mathcal{M}^\alpha) \to \mathcal{H}^2(\mathcal{M}^\alpha, \mathbb{Z})$ is surjective. Therefore, the map $d$ must be trivial, and the filtration of the K-groups, as shown in Fig. S1, gives the group extensions:

$$0 \to \mathcal{H}^0(\mathcal{M}^\alpha, \mathbb{Z}) \to K^0(\mathcal{M}^\alpha) \to \mathcal{H}^2(\mathcal{M}^\alpha, \mathbb{Z}) \to 0, \tag{S11}$$

and

$$0 \to \mathcal{H}^1(\mathcal{M}^\alpha, \mathbb{Z}) \to K^1(\mathcal{M}^\alpha) \to \mathcal{H}^3(\mathcal{M}^\alpha, \mathbb{Z}) \to 0, \tag{S12}$$

for $\alpha = 0, 5, \ldots, 9$.

On the other hand, for non-orientable $\mathcal{M}^\alpha$ with $\alpha = 1, \ldots, 4$, the chain complex in Eq. (S8) takes the form:

$$0 \to \mathbb{Z} \xrightarrow{d} \mathbb{Z}_2 \to 0. \tag{S13}$$

The map $d$ can either be trivial or map to the parity of an integer. If $d$ is trivial, there is no update in the spectral sequence, i.e., $E_4^{p,q} \cong E_3^{p,q}$. However, if $d$ maps to the parity of an integer, the $E_4$ page will be updated as shown in Fig. S3.

In both cases, regardless of whether $d$ is trivial or not, the group extension of $K^0(\mathcal{M}^\alpha)$ can be given by

$$0 \to \mathcal{H}^0(\mathcal{M}^\alpha, \mathbb{Z}) \to K^0(\mathcal{M}^\alpha) \to \mathcal{H}^2(\mathcal{M}^\alpha, \mathbb{Z}) \to 0, \tag{S14}$$

for $\alpha = 1, \ldots, 4$. This has the same form as Eq. (S11). Moreover, the extension of $K^1(\mathcal{M}^\alpha)$ can be given by

$$0 \to \mathcal{H}^1(\mathcal{M}^\alpha, \mathbb{Z}) \to K^1(\mathcal{M}^\alpha) \to \mathcal{H}^3(\mathcal{M}^\alpha, \mathbb{Z}) \to 0, \tag{S15}$$

while if $d$ is nontrivial, the extension can be expressed as

$$0 \to \mathcal{H}^1(\mathcal{M}^\alpha, \mathbb{Z}) \to K^1(\mathcal{M}^\alpha) \to 0 \to 0. \tag{S16}$$

In conclusion, for all ten platycosms, $K^0(\mathcal{M}^\alpha)$ can be computed using the group extension

$$0 \to \mathbb{Z} \to K^0(\mathcal{M}^\alpha) \to \mathcal{H}^2(\mathcal{M}^\alpha, \mathbb{Z}) \to 0, \tag{S17}$$

where we use $\mathcal{H}^0(\mathcal{M}^\alpha, \mathbb{Z}) \cong \mathbb{Z}$. For $K^1(\mathcal{M}^\alpha)$, when $\mathcal{M}^\alpha$ is orientable, it's group extension must be given by

$$0 \to \mathcal{H}^1(\mathcal{M}^\alpha, \mathbb{Z}) \to K^1(\mathcal{M}^\alpha) \to \mathbb{Z} \to 0. \tag{S18}$$

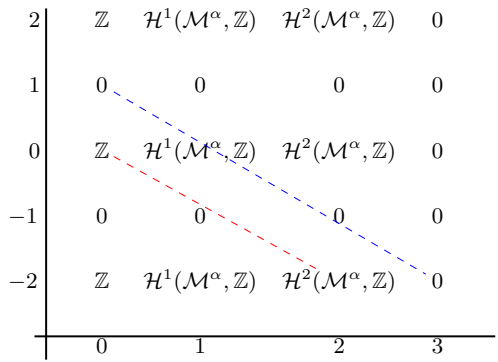

FIG. S3: $E_4$ pages of non-orientable $\mathcal{M}^\alpha$ with nontrivial map $d$ and $\alpha = 1, \ldots, 4$. The horizontal axis corresponds to $p$ and the vertical axis to $q$. The red dashed line represents filtration of $K^0(\mathcal{M}^\alpha)$ and the blue dashed line represents filtration of $K^1(\mathcal{M}^\alpha)$.

However, when $\mathcal{M}^\alpha$ is non-orientable, $K^1(\mathcal{M}^\alpha)$ can be analyzed by the group extension

$$0 \to \mathcal{H}^1(\mathcal{M}^\alpha, \mathbb{Z}) \to K^1(\mathcal{M}^\alpha) \to \mathbb{Z}_2 \to 0, \tag{S19}$$

or

$$0 \to \mathcal{H}^1(\mathcal{M}^\alpha, \mathbb{Z}) \to K^1(\mathcal{M}^\alpha) \to 0 \to 0. \tag{S20}$$

In the remainder of this section, we will compute the K groups by examining these group extensions on a case-by-case basis.

### Cubical torocosm $\mathcal{M}^0$

Observing from Tab. S1, the cohomology groups of $\mathcal{M}^0$ are

$$\mathcal{H}^1(\mathcal{M}^0, \mathbb{Z}) \cong \mathbb{Z}^3, \quad \mathcal{H}^2(\mathcal{M}^0, \mathbb{Z}) \cong \mathbb{Z}^3. \tag{S21}$$

The group extension of $K^0(\mathcal{M}^0)$ is given by

$$0 \to \mathbb{Z} \to K^0(\mathcal{M}^0) \to \mathbb{Z}^3 \to 0, \tag{S22}$$

and the group extension of $K^1(\mathcal{M}^0)$ can be written as

$$0 \to \mathbb{Z}^3 \to K^1(\mathcal{M}^0) \to \mathbb{Z} \to 0. \tag{S23}$$

The extensions can only be trivial, so we have $K^0(\mathcal{M}^0) \cong \mathbb{Z}^4$ and $K^1(\mathcal{M}^0) \cong \mathbb{Z}^4$. Furthermore, the reduced K group $\widetilde{K}(\mathcal{M}^0)$ is isomorphic to $\mathbb{Z}^3$.

### First amphicosm $\mathcal{M}^1$

Observing from Tab. S1, the cohomology groups of $\mathcal{M}^1$ are

$$\mathcal{H}^1(\mathcal{M}^1, \mathbb{Z}) \cong \mathbb{Z}^2, \quad \mathcal{H}^2(\mathcal{M}^1, \mathbb{Z}) \cong \mathbb{Z}_2 \oplus \mathbb{Z}. \tag{S24}$$

The group extension of $K^0(\mathcal{M}^1)$ is given by

$$0 \to \mathbb{Z} \to K^0(\mathcal{M}^1) \to \mathbb{Z}_2 \oplus \mathbb{Z} \to 0, \tag{S25}$$

and the group extension of $K^1(\mathcal{M}^1)$ can be written as

$$0 \to \mathbb{Z}^3 \to K^1(\mathcal{M}^1) \to \mathbb{Z}_2 \to 0, \tag{S26}$$

or
$$0 \to \mathbb{Z}^3 \to K^1(\mathcal{M}^1) \to 0 \to 0. \tag{S27}$$

The extension in Eq. (S25) admits two possibilities, so $K^0(\mathcal{M}^1)$ can be isomorphic to either $\mathbb{Z}^2 \oplus \mathbb{Z}_2$ or $\mathbb{Z}^2$. However, because of the condition that the determinant map from $\widetilde{K}$ to $\mathcal{H}^2$ is surjective, we conclude that $K^0(\mathcal{M}^1) \cong \mathbb{Z}^2 \oplus \mathbb{Z}_2$ and $\widetilde{K}(\mathcal{M}^1) \cong \mathbb{Z} \oplus \mathbb{Z}_2$.

Additionally, $K^1(\mathcal{M}^1)$ can be isomorphic to either $\mathbb{Z}^2 \oplus \mathbb{Z}_2$ or $\mathbb{Z}^2$.

### Second amphicosm $\mathcal{M}^2$

Observing from Tab. S1, the cohomology groups of $\mathcal{M}^2$ are
$$\mathcal{H}^1(\mathcal{M}^2, \mathbb{Z}) \cong \mathbb{Z}^2, \ \mathcal{H}^2(\mathcal{M}^2, \mathbb{Z}) \cong \mathbb{Z}. \tag{S28}$$

The group extension of $K^0(\mathcal{M}^2)$ is given by
$$0 \to \mathbb{Z} \to K^0(\mathcal{M}^2) \to \mathbb{Z} \to 0, \tag{S29}$$

and the group extension of $K^1(\mathcal{M}^2)$ can be written as
$$0 \to \mathbb{Z}^2 \to K^1(\mathcal{M}^2) \to \mathbb{Z}_2 \to 0, \tag{S30}$$

or
$$0 \to \mathbb{Z}^2 \to K^1(\mathcal{M}^2) \to 0 \to 0. \tag{S31}$$

The extension in Eq. (S29) must be trivial, so $K^0(\mathcal{M}^2) \cong \mathbb{Z}^2$ and $\widetilde{K}(\mathcal{M}^2) \cong \mathbb{Z}$.

In addition, $K^1(\mathcal{M}^2)$ can be isomorphic to $\mathbb{Z}^2 \oplus \mathbb{Z}_2$ or $\mathbb{Z}^2$.

### First amphidicosm $\mathcal{M}^3$

Observing from Tab. S1, the cohomology groups of $\mathcal{M}^3$ are
$$\mathcal{H}^1(\mathcal{M}^3, \mathbb{Z}) \cong \mathbb{Z}, \ \mathcal{H}^2(\mathcal{M}^3, \mathbb{Z}) \cong \mathbb{Z}_2^2. \tag{S32}$$

The group extension of $K^0(\mathcal{M}^3)$ is given by
$$0 \to \mathbb{Z} \to K^0(\mathcal{M}^3) \to \mathbb{Z}_2^2 \to 0, \tag{S33}$$

and the group extension of $K^1(\mathcal{M}^3)$ can be written as
$$0 \to \mathbb{Z} \to K^1(\mathcal{M}^3) \to \mathbb{Z}_2 \to 0, \tag{S34}$$

or
$$0 \to \mathbb{Z} \to K^1(\mathcal{M}^3) \to 0 \to 0. \tag{S35}$$

With extension in Eq. (S33), $K^0(\mathcal{M}^3)$ can be isomorphic to $\mathbb{Z} \oplus \mathbb{Z}_2^2$ or $\mathbb{Z} \oplus \mathbb{Z}_2$. Due to the restriction imposed by $\mathcal{H}^2(\mathcal{M}^3, \mathbb{Z})$, we conclude that $K^0(\mathcal{M}^3) \cong \mathbb{Z} \oplus \mathbb{Z}_2^2$ and $\widetilde{K}(\mathcal{M}^3) \cong \mathbb{Z}_2^2$.

In addition, $K^1(\mathcal{M}^3)$ can be isomorphic to $\mathbb{Z} \oplus \mathbb{Z}_2$ or $\mathbb{Z}$.

### Second amphidicosm $\mathcal{M}^4$

Observing from Tab. S1, the cohomology groups of $\mathcal{M}^4$ are
$$\mathcal{H}^1(\mathcal{M}^4, \mathbb{Z}) \cong \mathbb{Z}, \ \mathcal{H}^2(\mathcal{M}^4, \mathbb{Z}) \cong \mathbb{Z}_4. \tag{S36}$$

The group extension of $K^0(\mathcal{M}^4)$ is given by

$$0 \to \mathbb{Z} \to K^0(\mathcal{M}^4) \to \mathbb{Z}_4 \to 0, \tag{S37}$$

and the group extension of $K^1(\mathcal{M}^4)$ can be written as

$$0 \to \mathbb{Z} \to K^1(\mathcal{M}^4) \to \mathbb{Z}_2 \to 0, \tag{S38}$$

or

$$0 \to \mathbb{Z} \to K^1(\mathcal{M}^4) \to 0 \to 0. \tag{S39}$$

With extension in Eq. (S37), $K^0(\mathcal{M}^4)$ can be isomorphic to $\mathbb{Z}\oplus\mathbb{Z}_4$ or $\mathbb{Z}$. Due to the restriction imposed by $\mathcal{H}^2(\mathcal{M}^4, \mathbb{Z})$, we conclude that $K^0(\mathcal{M}^4) \cong \mathbb{Z} \oplus \mathbb{Z}_4$ and $\widetilde{K}(\mathcal{M}^4) \cong \mathbb{Z}_4$.

Additionally, $K^1(\mathcal{M}^4)$ can be isomorphic to $\mathbb{Z} \oplus \mathbb{Z}_2$ or $\mathbb{Z}$.

### Dicosm $\mathcal{M}^5$

Observing from Tab. S1, the cohomology groups of $\mathcal{M}^5$ are

$$\mathcal{H}^1(\mathcal{M}^5, \mathbb{Z}) \cong \mathbb{Z}, \quad \mathcal{H}^2(\mathcal{M}^5, \mathbb{Z}) \cong \mathbb{Z}_2^2 \oplus \mathbb{Z}. \tag{S40}$$

The group extension of $K^0(\mathcal{M}^5)$ is given by

$$0 \to \mathbb{Z} \to K^0(\mathcal{M}^5) \to \mathbb{Z}_2^2 \oplus \mathbb{Z} \to 0, \tag{S41}$$

and the group extension of $K^1(\mathcal{M}^5)$ is given by

$$0 \to \mathbb{Z} \to K^1(\mathcal{M}^5) \to \mathbb{Z} \to 0. \tag{S42}$$

From the extension in Eq. (S41), $K^0(\mathcal{M}^5)$ can be isomorphic to either $\mathbb{Z}^2 \oplus \mathbb{Z}_2^2$ or $\mathbb{Z}^2 \oplus \mathbb{Z}_2$. However, due to the restriction imposed by $\mathcal{H}^2(\mathcal{M}^5, \mathbb{Z})$, we conclude that $K^0(\mathcal{M}^5) \cong \mathbb{Z}^2 \oplus \mathbb{Z}_2^2$ and $\widetilde{K}(\mathcal{M}^5) \cong \mathbb{Z} \oplus \mathbb{Z}_2^2$.

Additionally, the extension in Eq. (S42) for $K^1(\mathcal{M}^5)$ must be trivial, so we find $K^1(\mathcal{M}^5) \cong \mathbb{Z}^2$.

### Tricosm $\mathcal{M}^6$

Observing from Tab. S1, the cohomology groups of $\mathcal{M}^6$ are

$$\mathcal{H}^1(\mathcal{M}^6, \mathbb{Z}) \cong \mathbb{Z}, \ \mathcal{H}^2(\mathcal{M}^6, \mathbb{Z}) \cong \mathbb{Z}_3 \oplus \mathbb{Z}. \tag{S43}$$

The group extension of $K^0(\mathcal{M}^6)$ is given by

$$0 \to \mathbb{Z} \to K^0(\mathcal{M}^6) \to \mathbb{Z}_3 \oplus \mathbb{Z} \to 0, \tag{S44}$$

and the group extension of $K^1(\mathcal{M}^6)$ is given by

$$0 \to \mathbb{Z} \to K^1(\mathcal{M}^6) \to \mathbb{Z} \to 0, \tag{S45}$$

With extension in Eq. (S44), $K^0(\mathcal{M}^6)$ can be isomorphic to $\mathbb{Z}^2 \oplus \mathbb{Z}_3$ or $\mathbb{Z}^2$. Due to the restriction imposed by $\mathcal{H}^2(\mathcal{M}^6, \mathbb{Z})$, we conclude that $K^0(\mathcal{M}^6) \cong \mathbb{Z}^2 \oplus \mathbb{Z}_3$ and $\widetilde{K}(\mathcal{M}^6) \cong \mathbb{Z} \oplus \mathbb{Z}_3$.

Similarly to the previous discussion, we have $K^1(\mathcal{M}^6) \cong \mathbb{Z}^2$.

### Tetracosm $\mathcal{M}^7$

Observing from Tab. S1, the cohomology groups of $\mathcal{M}^7$ are

$$\mathcal{H}^1(\mathcal{M}^7, \mathbb{Z}) \cong \mathbb{Z}, \ \mathcal{H}^2(\mathcal{M}^7, \mathbb{Z}) \cong \mathbb{Z}_2 \oplus \mathbb{Z}. \tag{S46}$$

The group extension of $K^0(\mathcal{M}^7)$ is given by

$$0 \to \mathbb{Z} \to K^0(\mathcal{M}^7) \to \mathbb{Z}_2 \oplus \mathbb{Z} \to 0, \tag{S47}$$

and the group extension of $K^1(\mathcal{M}^7)$ is given by

$$0 \to \mathbb{Z} \to K^1(\mathcal{M}^7) \to \mathbb{Z} \to 0, \tag{S48}$$

With extension in Eq. (S47), $K^0(\mathcal{M}^7)$ can be isomorphic to $\mathbb{Z}^2 \oplus \mathbb{Z}_2$ or $\mathbb{Z} \oplus \mathbb{Z}_2$. Due to the restriction imposed by $\mathcal{H}^2(\mathcal{M}^7, \mathbb{Z})$, we conclude that $K^0(\mathcal{M}^7) \cong \mathbb{Z} \oplus \mathbb{Z}_3$ and $\widetilde{K}(\mathcal{M}^7) \cong \mathbb{Z} \oplus \mathbb{Z}_2$.

Similarly to the previous discussion, we have $K^1(\mathcal{M}^7) \cong \mathbb{Z}^2$.

### Hexacosm $\mathcal{M}^8$

Observing from Tab. S1, the cohomology groups of $\mathcal{M}^8$ are

$$\mathcal{H}^1(\mathcal{M}^8, \mathbb{Z}) \cong \mathbb{Z}, \ \mathcal{H}^2(\mathcal{M}^8, \mathbb{Z}) \cong \mathbb{Z}. \tag{S49}$$

The group extension of $K^0(\mathcal{M}^8)$ is given by

$$0 \to \mathbb{Z} \to K^0(\mathcal{M}^8) \to \mathbb{Z} \to 0, \tag{S50}$$

and the group extension of $K^1(\mathcal{M}^8)$ is given by

$$0 \to \mathbb{Z} \to K^1(\mathcal{M}^8) \to \mathbb{Z} \to 0, \tag{S51}$$

The extension in Eq. (S50) must be trivial and therefore $K^0(\mathcal{M}^8) \cong \mathbb{Z}^2$ and $\widetilde{K}(\mathcal{M}^8) \cong \mathbb{Z}$.

Similarly to the previous discussion, we have $K^1(\mathcal{M}^8) \cong \mathbb{Z}^2$.

### Didicosm $\mathcal{M}^9$

Observing from Tab. S1, the cohomology groups of $\mathcal{M}^9$ are

$$\mathcal{H}^1(\mathcal{M}^9, \mathbb{Z}) \cong \mathbb{Z}, \ \mathcal{H}^2(\mathcal{M}^9, \mathbb{Z}) \cong \mathbb{Z}_4^2. \tag{S52}$$

The group extension of $K^0(\mathcal{M}^9)$ is given by

$$0 \to \mathbb{Z} \to K^0(\mathcal{M}^9) \to \mathbb{Z}_4^2 \to 0, \tag{S53}$$

and the group extension of $K^1(\mathcal{M}^9)$ is given by

$$0 \to 0 \to K^1(\mathcal{M}^9) \to \mathbb{Z} \to 0, \tag{S54}$$

With extension in Eq. (S53), $K^0(\mathcal{M}^9)$ can be isomorphic to $\mathbb{Z} \oplus \mathbb{Z}_4^2$ or $\mathbb{Z} \oplus \mathbb{Z}_4$. Due to the restriction imposed by $\mathcal{H}^2(\mathcal{M}^9, \mathbb{Z})$, we conclude that $K^0(\mathcal{M}^9) \cong \mathbb{Z} \oplus \mathbb{Z}_4^2$ and $\widetilde{K}(\mathcal{M}^9) \cong \mathbb{Z}_4^2$.

The extension in Eq. (S54) of $K^1(\mathcal{M}^9)$ is trivial, so we have $K^1(\mathcal{M}^9) \cong \mathbb{Z}$.

## III. The complete set of topological invariants

In this section, we present concrete constructions of topological invariants that characterize topological classifications over ten Brillouin platycosms.

### Cubical torocosm $\mathcal{M}^0$

The momentum space Bieberbach group corresponding to the Brillouin cubical torocosm $\mathcal{M}^0$ is $P1$, which exhibits only translation symmetries along three directions. The cubical torocosm is synonymous with the 3D torus, and the classification $\mathbb{Z}^3$ is a well-known result, characterized by the Chern numbers over three distinct 2D sub-tori. We can denote the Chern number as $\text{Ch}^{k_x}$ (or $\text{Ch}^{k_y}, \text{Ch}^{k_z}$) to represent the one defined in the $k_y$-$k_z$ (or $k_x$-$k_z$, $k_x$-$k_y$) plane for fixed values of $k_x$ (or $k_y$, $k_z$).

### First amphicosm $\mathcal{M}^1$

The momentum space Bieberbach group corresponding to the Brillouin first amphicosm is $Pc$. In this context, we adopt the convention that the point group $D_1 = \{E, \mathcal{G}_x\}$ of $Pc$ acts on momentum space as follows:

$$\mathcal{G}_x : (k_x, k_y, k_z) \to (-k_x, k_y, k_z + \frac{1}{2}G_z). \tag{S55}$$

The reduced K group $\widetilde{K}(\mathcal{M}^0) \cong \mathbb{Z} \oplus \mathbb{Z}_2$ reflects the presence of a Chern number and a $\mathbb{Z}_2$-invariant. These invariants can be expressed as:

$$\begin{aligned}
\text{Ch} &= \frac{1}{2\pi} \int_M f, \\
\nu^{(2)} &= \frac{1}{2\pi} \int_X f - \frac{2}{\pi} \int_{S^+} a \mod 2.
\end{aligned} \tag{S56}$$

Here, the 2D sub-manifolds $M$, $X$, and the oriented segment $S^+$ are illustrated in Fig. S4(a).

On the planes $k_x = 0$ or $k_x = \pi$, the glide reflection $\mathcal{G}_x$ acts as a half-translation along the $k_z$-direction. Therefore, the sub-manifold $X$ is a 2D torus, and a Chern number can be defined over it. The Chern number for the plane defined by $k_y$-$k_z$ at an arbitrary $k_x$ is 2Ch. Moreover, due to the glide reflection symmetry $\mathcal{G}_x$, the sub-manifold $X$ has the topology of a 2D Klein bottle. The $\mathbb{Z}_2$-invariant $\nu^{(2)}$ is defined over $X$ as described in the main text.

Furthermore, it is straightforward to verify that the Chern numbers vanish in the other two directions. Specifically, on the $k_x$-$k_y$ and $k_x$-$k_z$ planes, the $k_x$-coordinate is reversed by the glide reflection $\mathcal{G}_x$.

### Second amphicosm $\mathcal{M}^2$

The momentum space Bieberbach group corresponding to the second amphicosm is $Cc$. We assume that the point group $D_1 = \{E, \mathcal{G}\}$ acts on momentum space as follows:

$$\mathcal{G} : (k_x, k_y, k_z) \to (k_y, k_x, k_z + \frac{1}{2}G_z). \tag{S57}$$

The reduced K group $\widetilde{K}(\mathcal{M}^2)$ is isomorphic to $\mathbb{Z}$, and the only topological invariant is the Chern number, which is given by:

$$\text{Ch} = \frac{1}{2\pi} \int_M f, \tag{S58}$$

where $M$ is shown in Fig. S4(b).

Similar to the case of $Pc$, the Chern number can be defined on half of the plane $k_y = k_x$. Furthermore, the Chern number on the $k_x$-$k_y$ plane vanishes. While it appears that we could define a $\mathbb{Z}_2$-invariant on $X$, since $X$ is also with 2D Klein bottle topology under the action $\mathcal{G}$, we can prove that the value of this invariant is dependent on the Chern number Ch. The detailed proof will be presented in the next section.

### First amphidicosm $\mathcal{M}^3$

The momentum space Bieberbach group corresponding to the first amphidicosm is $Pca2_1$. We assume the action of the point group $D_2 = \{E, \mathcal{G}_x, \mathcal{G}_y, \mathcal{G}_x\mathcal{G}_y\}$ on momentum space is given by:

$$\begin{aligned}
\mathcal{G}_x &: (k_x, k_y, k_z) \to (-k_x, k_y, k_z + \frac{1}{2}G_z), \\
\mathcal{G}_y &: (k_x, k_y, k_z) \to (k_x + \frac{1}{2}G_x, -k_y, k_z).
\end{aligned} \tag{S59}$$

The reduced K group $\widetilde{K}(\mathcal{M}^3)$ is isomorphic to $\mathbb{Z}_2^2$. The two $\mathbb{Z}_2$-invariants can be defined as:

$$\begin{aligned}
\nu_1^{(2)} &= \frac{1}{2\pi} \int_{X_1} f - \frac{1}{\pi} \int_{S_1^+} a \mod 2, \\
\nu_2^{(2)} &= \frac{1}{2\pi} \int_{X_2} f - \frac{1}{\pi} \int_{S_2^+} a \mod 2.
\end{aligned} \tag{S60}$$

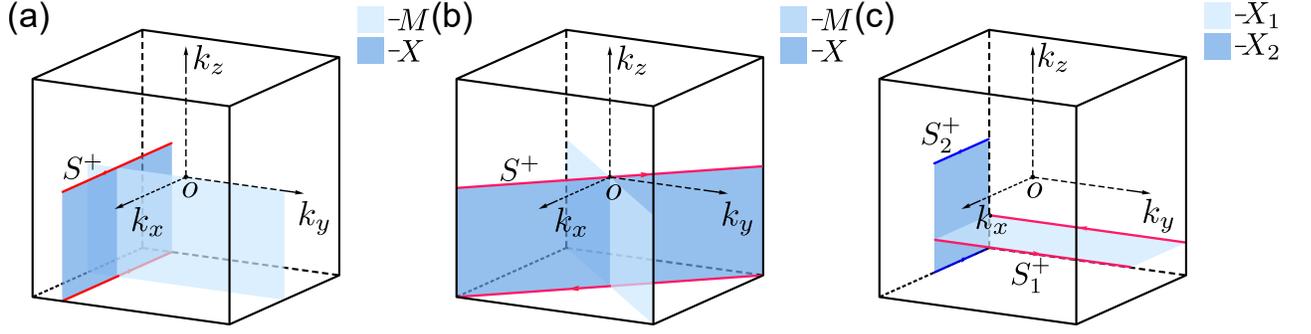

FIG. S4: Brillouin zone of (a) $k$-space $Pc$, (b) $k$-space $Cc$, (c) $k$-space $Pca2_1$.

Here, the 2D sub-manifolds $X_{1,2}$ and the oriented segments $S^+_{1,2}$ are illustrated in Fig. S4(c). The sub-manifold $X_1$ has the topology of a Klein bottle under the action of $\mathcal{G}_y$, so the $\mathbb{Z}_2$-invariant $\nu^{(2)}_1$ is well-defined.

On the other hand, we can also define a $\mathbb{Z}_2$-invariant over half of the $k_x$-$k_z$ plane, since the glide reflection $\mathcal{G}_x$ gives it a Klein bottle topology. However, for the planes where $k_y = 0$ or $k_y = \pi$, the other glide reflection $\mathcal{G}_y$ acts as a half-translation symmetry along the $k_x$ direction. It assumes the periodic boundary conditions along the $k_x$ direction as imposed by reciprocal translation along the $k_x$ direction. This leads to the definition of $\nu^{(2)}_2$. Meanwhile, the $\mathbb{Z}_2$ invariant defined on half of the plane equals to $2\nu^{(2)}_2$ and therefore must be trivial.

### Second amphidicosm $\mathcal{M}^4$

The momentum space Bieberbach group corresponding to the second amphidicosm is $Pna2_1$. We assume the action of the point group $D_2 = \{E, \mathcal{G}_x, \mathcal{G}_y, \mathcal{G}_x\mathcal{G}_y\}$ on momentum space is given by:

$$\begin{aligned}
\mathcal{G}_x &: (k_x, k_y, k_z) \to (-k_x, k_y + \frac{1}{2}\boldsymbol{G}_y, k_z + \frac{1}{2}\boldsymbol{G}_z), \\
\mathcal{G}_y &: (k_x, k_y, k_z) \to (k_x + \frac{1}{2}\boldsymbol{G}_x, -k_y, k_z).
\end{aligned} \tag{S61}$$

The reduced K group $\widetilde{K}(\mathcal{M}^4)$ is isomorphic to $\mathbb{Z}_4$. The $\mathbb{Z}_4$-invariant is defined over the plane $k_y = k_z$, as follows:

$$\nu^{(4)} = \frac{1}{2\pi}\int_X f - \frac{2}{\pi}\int_{S^+} a \mod 4. \tag{S62}$$

Here, the 2D sub-manifold $X$ and the oriented segment $S^+$ are illustrated in Fig. S5(a). This is the case that has been introduced in the main text. The glide reflection symmetry $\mathcal{G}_y$ acts on the planes $k_y = 0$ or $k_y = \pi$ as a half-translation along the $k_x$-direction, assuming periodic boundary conditions along the $k_x$-direction, as imposed by the reciprocal translation along $k_x$. The two glide reflection symmetries $\mathcal{G}_x$ and $\mathcal{G}_y$ act nontrivially on the boundary $\partial X$ of $X$, specifically on the two $k_x$-edges. The fundamental domain $\partial X/G$ is half of an edge with an orientation, and we may choose it as the oriented segment $S^+$, as indicated in Fig. S5(a). Thus, the topological invariants is given by Eq. (S61).

### Dicosm $\mathcal{M}^5$

The momentum space Bieberbach group corresponding to the Dicosm is $P2_1$. We assume the action of the point group $C_2 = \{E, \mathcal{R}_2\}$ on momentum space is given by:

$$\mathcal{R}_2 : (k_x, k_y, k_z) \to (-k_x, -k_y, k_z + \frac{1}{2}\boldsymbol{G}_z). \tag{S63}$$

The reduced K group $\widetilde{K}(\mathcal{M}^5) \cong \mathbb{Z} \oplus \mathbb{Z}_2^2$ reflects the presence of a Chern number and two $\mathbb{Z}_2$-invariants. These invariants can be expressed as:

$$\begin{aligned} \text{Ch} &= \frac{1}{2\pi} \int_M f, \\ \nu_1^{(2)} &= \frac{1}{2\pi} \int_{X_1} f - \frac{1}{\pi} \int_{S_1^+} a \mod 2, \\ \nu_2^{(2)} &= \frac{1}{2\pi} \int_{X_2} f - \frac{1}{\pi} \int_{S_2^+} a \mod 2. \end{aligned} \quad (S64)$$

Here, the 2D sub-manifolds $X_{1,2}, M$ and the oriented segments $S_{1,2}^+$ are illustrated in Fig. S5(b). The two boundaries $S_1^\pm$ of $X_1$ are oriented segments along the $k_x$-direction, which are transformed onto each other by the screw rotation symmetry $\mathcal{R}_2$. Thus, $X_1$ has the topology of a Klein bottle, and therefore $\nu_1^{(2)}$ is well-defined. A similar argument holds for $X_2$ and $\nu_2^{(2)}$.

There are several alternative ways to define the $\mathbb{Z}_2$ invariants, all of which are equivalent to the definition presented here. These alternative definitions will be introduced in the next section.

Moreover, since $\mathcal{R}_2$ does not change the orientation of the $k_x$-$k_y$ plane, a Chern number is well-defined over this plane.

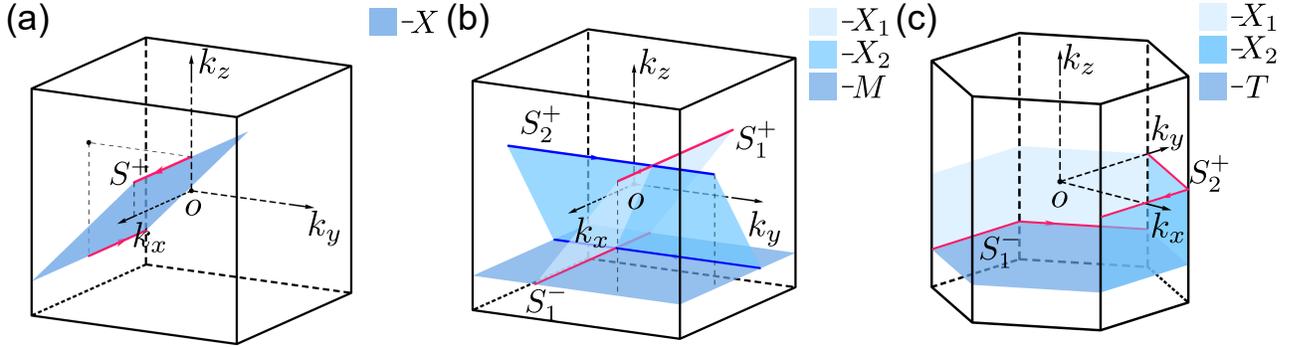

FIG. S5: Brillouin zone of (a) $k$-space $Pna2_1$, (b) $k$-spae $P2_1$, (c) $k$-space $P3_1$.

### Tricosm $\mathcal{M}^6$

The momentum space Bieberbach group corresponding to the tricosm is $P3_1$. We assume the action of the point group $C_3 = \{E, \mathcal{R}_3, \mathcal{R}_3^2\}$ on momentum space is given by:

$$\mathcal{R}_3 : (k_x, k_y, k_z) \to \left(-\frac{\sqrt{3}}{2} k_x + \frac{1}{2} k_y, -\frac{1}{2} k_x - \frac{\sqrt{3}}{2} k_y, k_z + \frac{1}{3} \boldsymbol{G}_z\right). \quad (S65)$$

The reduced K group $\widetilde{K}(\mathcal{M}^6)$ of tricosm is isomorphic to $\mathbb{Z}_3 \oplus \mathbb{Z}$. The invariants can be defined as:

$$\begin{aligned} \nu^{(3)} &= \frac{1}{2\pi} \int_{X_1 - X_2} f + \frac{3}{2\pi} \int_{S_1^-} a \mod 3, \\ \text{Ch} &= \frac{1}{2\pi} \int_T f. \end{aligned} \quad (S66)$$

Here, the cylinders $X_1, X_2$ and the oriented segment $S_1^-$ are illustrated in Fig. S5(c). This is the case introduced in the main text. The boundary $\partial X_i$ consists of two oriented components $S_i^\pm$ with $i = 1, 2, 3$, and the screw rotation maps $S_i^+$ to $S_{i+1}^+$ with $i \mod 3$, i.e.,

$$\partial X_i = S_i^+ - S_i^-, \quad \mathcal{R}_3 S_i^- = S_{i+1}^+. \quad (S67)$$

11Additionally, the boundary of the hexagon $T$ on the $k_x$-$k_y$ plane is given by:

$$\partial T = S_1^- + S_2^- + S_3^-. \tag{S68}$$

Note that the hexagon is topologically a torus under reciprocal lattice translations. We can identify the segments $S_i^\pm$ related by the screw rotation and derive the following:

$$\partial(X_1 - X_2 - T) = -3S_1^-. \tag{S69}$$

Thus, we can define $\nu^{(3)}$ in the form shown in Eq. (S66). Additionally, there exists a Chern number over the $k_x$-$k_y$ plane.

### Tetracosm $\mathcal{M}^7$

The momentum space Bieberbach group corresponding to the tetracosm is $P4_1$. We assume the action of the point group $C_4 = \{E, \mathcal{R}_4, \mathcal{R}_4^2, \mathcal{R}_4^3\}$ on momentum space is given by:

$$\mathcal{R}_4 : (k_x, k_y, k_z) \to \left(k_y, -k_x, k_z + \frac{1}{4}\boldsymbol{G}_z\right). \tag{S70}$$

The reduced K group $\widetilde{K}(\mathcal{M}^7)$ is isomorphic to $\mathbb{Z}_2 \oplus \mathbb{Z}$. These invariants can be defined as:

$$\begin{aligned}\nu^{(2)} &= \frac{1}{2\pi}\int_X f - \frac{1}{\pi}\int_{S^+} a \mod 2, \\ \text{Ch} &= \frac{1}{2\pi}\int_M f.\end{aligned} \tag{S71}$$

Here, the sub-manifolds $X, M$ and the oriented segment $S^+$ are illustrated in Fig. S6(a). The $k$-space $P4_1$ contains $P2_1$ as a subgroup. Therefore, the invariants defined for $P2_1$ also hold for $P4_1$.

Additionally, the two $\mathbb{Z}_2$-invariants $\nu_1^{(2)}$ and $\nu_2^{(2)}$ must be the same due to the additional 4-fold screw rotation symmetry $\mathcal{R}_4$. This explains why there is only one independent $\mathbb{Z}_2$-invariant $\nu^{(2)}$ in this case.

### Hexacosm $\mathcal{M}^8$

The momentum space Bieberbach group corresponding to the Hexacosm is $P6_1$. We assume the action of the point group $C_6 = \{E, \mathcal{R}_6, \mathcal{R}_6^2, \mathcal{R}_6^3, \mathcal{R}_6^4, \mathcal{R}_6^5\}$ on momentum space is given by:

$$\mathcal{R}_6 : (k_x, k_y, k_z) \to \left(\frac{1}{2}k_x + \frac{\sqrt{3}}{2}k_y, -\frac{\sqrt{3}}{2}k_x + \frac{1}{2}k_y, k_z + \frac{1}{6}\boldsymbol{G}_z\right). \tag{S72}$$

The reduced K group $\widetilde{K}(\mathcal{M}^8)$ is isomorphic to $\mathbb{Z}$. The only invariant is the Chern number, which can be expressed as:

$$\text{Ch} = \frac{1}{2\pi}\int_M f. \tag{S73}$$

Here, the sub-manifold $M$ is illustrated in Fig. S6(b). Since $P3_1$ is a subgroup of $P6_1$, both $\nu^{(3)}$ and Ch are well defined in this case. However, the 6-fold screw rotation symmetry requires the $\mathbb{Z}_3$-invariant to be trivial. The detailed proof can be found in the next section.

### Didicosm $\mathcal{M}^9$

The momentum space Bieberbach group corresponding to the Didicosm is $P2_12_12_1$. We assume the action of its point group $D_2 = \{E, \mathcal{R}_y, \mathcal{R}_z, \mathcal{R}_y\mathcal{R}_z\}$ on momentum space is given by:

$$\begin{aligned}\mathcal{R}_y &: (k_x, k_y, k_z) \to \left(-k_x + \frac{1}{2}\boldsymbol{G}_x, k_y + \frac{1}{2}\boldsymbol{G}_y, -k_z\right), \\ \mathcal{R}_z &: (k_x, k_y, k_z) \to \left(-k_x, -k_y + \frac{1}{2}\boldsymbol{G}_y, k_z + \frac{1}{2}\boldsymbol{G}_z\right).\end{aligned} \tag{S74}$$

11

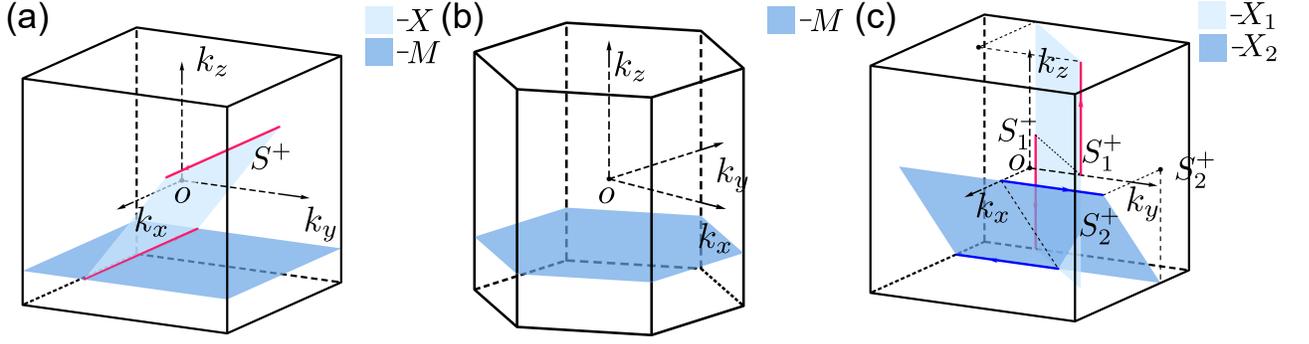

FIG. S6: Brillouin zone of (a) $k$-space $P4_1$, (b) $k$-space $P6_1$, (c) $k$-space $P2_12_12_1$.

The reduced K group $\widetilde{K}(\mathcal{M}^9)$ is isomorphic to $\mathbb{Z}_4^2$. The two $\mathbb{Z}_4$-invariants can be expressed as:

$$\begin{aligned}
\nu_1^{(4)} &= \frac{1}{2\pi}\int_{X_1} f - \frac{2}{\pi}\int_{S_1^+} a \mod 4, \\
\nu_2^{(4)} &= \frac{1}{2\pi}\int_{X_2} f - \frac{2}{\pi}\int_{S_2^+} a \mod 4.
\end{aligned} \quad (S75)$$

Here, the sub-manifolds $X_1, X_2$ and the oriented segments $S_1^\pm, S_2^\pm$ are illustrated in Fig. S6(c). We take the boundaries of $X_1$ to be fixed in the planes $k_x = 0$ and $k_x = -\pi$ respectively. Thus, $X_1$ has the topology of a Klein bottle imposed by the screw rotation symmetry $\mathcal{R}_y$, and we can check that the oriented segment $S_1^+$ generates the boundary $\partial X_1$ under the action of $D_2$. Therefore, the invariant $\nu_1^{(4)}$ is of a similar form to the one for the $k$-space $Pna2_1$.

A similar argument holds for $X_2$ and $\nu_2^{(4)}$, since it also has the topology of a Klein bottle imposed by the screw rotation symmetry $\mathcal{R}_z$. And the two boundaries of $X_2$ are fixed in the planes $k_z = 0$ and $k_z = -\pi$ respectively.

## IV. Equivalence relations between topological invariants

In the previous section, we presented the construction of topological invariants for each classification. Obviously, there may be several different ways to define equivalent topological invariants. In this section, we take three examples to illustrate the equivalent relationships between topological invariants. This will include the cases discussed earlier.

### Example 1: Second amphicosm $\mathcal{M}^2$

First, reviewing the case of $k$-space $Cc$, it appears that we can define a $\mathbb{Z}_2$ invariant over $X$, as illustrated in Fig. S7(a). Through a continuous deformation, indicated by the dashed lines in Fig. S7(a), the sub-manifold $X$ can be deformed into $X_1 + X_2$, as shown in Fig. S7(b).

Thus, we can also define the $\mathbb{Z}_2$ invariant as:

$$\nu^{(2)} = \frac{1}{2\pi}\int_{X_1+X_2} f - \frac{1}{\pi}\int_{S_1^+ + S_2^+} a \mod 2. \quad (S76)$$

Here, the oriented segment $S_1^+ + S_2^+$ serves as the fundamental domain $\partial(X_1 + X_2)/B^2$. Next, we focus on the sub-manifold $X_1 + X_2 + M$, considering the periodicity of reciprocal space:

$$\frac{1}{2\pi}\int_{X_1+X_2+M} f - \frac{1}{\pi}\int_{S_1^+ + S_2^+} a = 0 \mod 2. \quad (S77)$$

A straightforward derivation then gives:

$$\text{Ch} = \frac{1}{2\pi}\int_M f = -\nu^{(2)} \mod 2. \quad (S78)$$

In conclusion, $\nu^{(2)}$ here is equivalent to the parity of the Chern number.

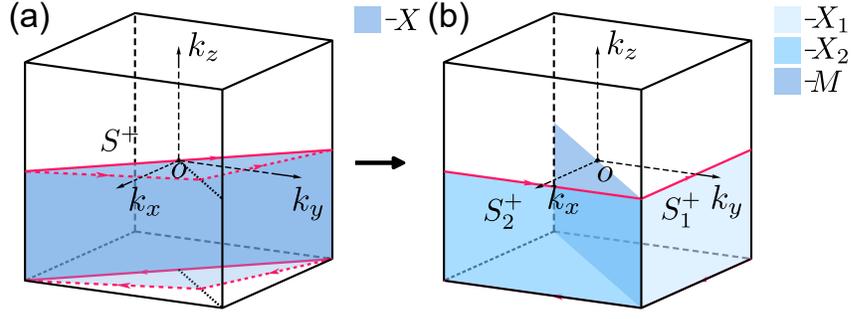

FIG. S7: (a) and (b) show two equivalent ways to define $\nu^{(2)}$ over $\mathcal{M}^2$.

### Example 2: Dicosm $\mathcal{M}^5$

The second example is the $k$-space $P2_1$. Consider the two $k_x$-$k_z$ planes with $k_y = 0$ and $k_y = \pi$. As illustrated in Fig. S8(a), the two boundaries $S_1^\pm(S_2^\pm)$ of $X_1(X_2)$ are transformed into each other by a 2-fold screw rotation symmetry $\mathcal{R}_2$. Therefore, we can define two $\mathbb{Z}_2$ invariants, expressed as:

$$\nu_0^{(2)} = \frac{1}{2\pi} \int_{X_1} f - \frac{1}{\pi} \int_{S_1^+} a,$$
$$\nu_\pi^{(2)} = \frac{1}{2\pi} \int_{X_2} f - \frac{1}{\pi} \int_{S_2^+} a. \tag{S79}$$

By continuously deforming $X_1$ along $k_y$ while keeping its two boundaries related by the screw rotation symmetry, we eventually deform $X_1^+$ into $X_1'$, as shown in Fig. S8(b). It is evident that $\nu_0^{(2)}$ is just $\nu_1^{(2)}$ as defined in Eq. (S64).

The sub-manifold $X_1' + X_2 + M$ is periodic in $k_z$, and the integration of flux over this sub-manifold gives a value that is a multiple of $2\pi$. We can then derive:

$$\nu_0^{(2)} + \nu_\pi^{(2)} = \frac{1}{2\pi} \int_{X_1+X_2} f = -\frac{1}{2\pi} \int_M f \mod 2. \tag{S80}$$

This shows that $\nu_0^{(2)}$ and $\nu_\pi^{(2)}$ are equal if the Chern number over $M$ is even, and unequal if the Chern number is odd. A similar argument holds for the $\mathbb{Z}_2$ invariants defined on the $k_y$-$k_z$ planes with $k_x = 0$ and $k_x = \pi$, which corresponds to $\nu_1^{(2)}$ as defined in Eq. (S64). Therefore, there are only two independent $\mathbb{Z}_2$ invariants and one independent Chern number in this case.

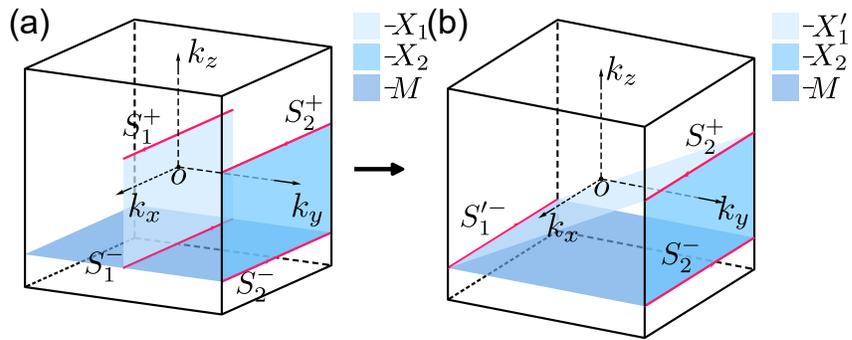

FIG. S8: (a) and (b) show Brillouin zone for $k$-space $P2_1$.

### Example 3: Tricosm $\mathcal{M}^6$ and Hexacosm $\mathcal{M}^8$

One definition of the $\mathbb{Z}_3$ invariant for $k$-space $P3_1$ is:

$$\nu^{(3)} = \frac{1}{2\pi}\int_{X_1-X_2} f + \frac{3}{2\pi}\int_{S_1^-} a \mod 3, \quad (S81)$$

where the cylinders $X_i$ and oriented segments $S_i^\pm$ with $i=1,2,3$ are shown in Fig. S9(a). However, as illustrated in Fig. S9(b), a similar but different $\nu^{(3)'}$ can be defined as:

$$\nu^{(3)'} = \frac{1}{2\pi}\int_{X_1'-X_2'} f + \frac{3}{2\pi}\int_{S_1'^-} a \mod 3, \quad (S82)$$

where $X_i'$ and $S_i'^\pm$ are obtained by rotating $X_i$ and $S_i^\pm$ by 60 degrees. Due to the periodic boundary conditions of reciprocal space, we can derive that:

$$X_1' - X_2' = X_1 - X_3, \quad S_1'^- = -S_3^-. \quad (S83)$$

If we add $\nu^{(3)}$ and $\nu^{(3)'}$, we get:

$$\begin{aligned}
\nu^{(3)} + \nu^{(3)'} &= \frac{1}{2\pi}\int_{2X_1-X_2-X_3} f + \frac{3}{2\pi}\int_{S_1^- - S_3^-} a \mod 3 \\
&= \frac{1}{2\pi}\int_{3X_1} f + \frac{3}{2\pi}\int_{S_1^- - S_1^+} a \mod 3 \\
&= 3\left(\frac{1}{2\pi}\int_{X_1} f - \frac{1}{2\pi}\int_{S_1^+ - S_1^-} a\right) = 0 \mod 3.
\end{aligned} \quad (S84)$$

Here, we identify $S_i^\pm$ related by the 3-fold screw rotation. Thus, there is only one independent $\mathbb{Z}_3$ invariant in this case.

Furthermore, $\nu^{(3)}$ and $\nu^{(3)'}$ are also well-defined if the $k$-space $P6_1$ symmetry holds. However, the 6-fold screw rotation symmetry requires that $\nu^{(3)} = \nu^{(3)'} \mod 3$. Along with the identity $\nu^{(3)} + \nu^{(3)'} = 0 \mod 3$, this means that there are no nontrivial $\mathbb{Z}_3$ invariants if the $k$-space $P6_1$ symmetry is present.

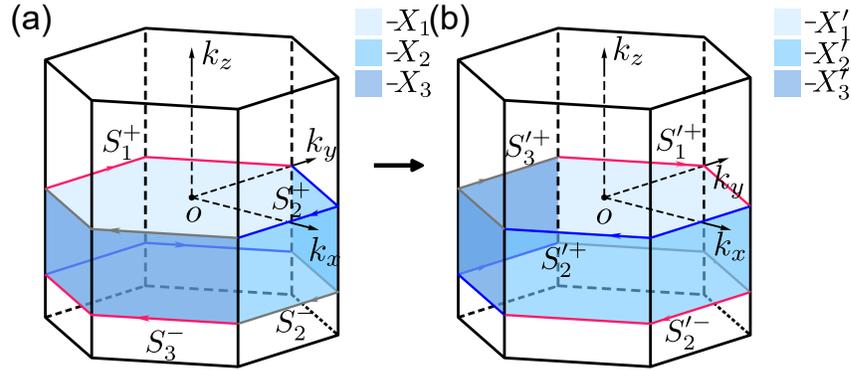

FIG. S9: (a) and (b) show Brillouin zone for $k$-space $P3_1$.

## V. Two-band Dirac models

We now consider a series of two-band Bloch Hamiltonians of the form $H^\alpha = \boldsymbol{d}^\alpha(\boldsymbol{k})\cdot\boldsymbol{\sigma}$ which are subject to corresponding Bieberbach groups. These toy models serve as straightforward examples for investigating topological phases over Brillouin platycosms.

### First amphicosm $\mathcal{M}^1$

The Hamiltonian with $k$-space $Pc$ symmetry must satisfy, according to our conventions, the relation

$$H^1(k_x, k_y, k_z) = U_{\mathcal{G}_x} H^1(-k_x, k_y, k_z + \pi) U_{\mathcal{G}_x}^\dagger. \tag{S85}$$

As a concrete example, we take

$$\begin{aligned}
d_x^1(\boldsymbol{k}) &= T_{xy} \cos(k_x) \sin(k_y) + T_{xz} \sin(k_x) \cos(k_z), \\
d_y^1(\boldsymbol{k}) &= T_{xy} \cos(k_x) \cos(k_y) + T_{xz} \sin(k_x) \sin(k_z), \\
d_z^1(\boldsymbol{k}) &= M + T_z \cos(2k_z) + T_x \cos(k_x).
\end{aligned} \tag{S86}$$

Here, $U_{\mathcal{G}_x}$ is an identity matrix. In main text Fig.1 (a), we plot a constant energy surface of this model, where parameters are taken as $E = -1.3, M = 0.5, T_x = T_{xy} = T_{xz} = 1, T_z = 0.5$.

### Second amphicosm $\mathcal{M}^2$

The Hamiltonian with $k$-space $Cc$ symmetry must satisfy

$$H^2(k_x, k_y, k_z) = U_{\mathcal{G}} H^2(k_y, k_x, k_z + \pi) U_{\mathcal{G}}^\dagger. \tag{S87}$$

As a concrete example, we take

$$\begin{aligned}
d_x^2(\boldsymbol{k}) &= T \cos(k_x) \cos(k_z) - T \cos(k_y) \sin(k_z), \\
d_y^2(\boldsymbol{k}) &= T \sin(k_x - k_y) \sin(k_z) + T_{xy} \sin(k_x) \sin(k_y), \\
d_z^2(\boldsymbol{k}) &= M + T_z \cos(2k_z).
\end{aligned} \tag{S88}$$

Here, $U_{\mathcal{G}_x}$ is an identity matrix. In main text Fig.1 (b), we plot a constant energy surface of this model, where parameters are taken as $E = -1.3, M = 0.5, T = 1, T_z = 0.5$.

### First amphidicosm $\mathcal{M}^3$

The Hamiltonian with $k$-space $Pca2_1$ symmetry must satisfy

$$\begin{aligned}
H^3(k_x, k_y, k_z) &= U_{\mathcal{G}_x} H^3(-k_x, k_y, k_z + \pi) U_{\mathcal{G}_x}^\dagger, \\
H^3(k_x, k_y, k_z) &= U_{\mathcal{G}_y} H^3(k_x + \pi, -k_y, k_z) U_{\mathcal{G}_y}^\dagger,
\end{aligned} \tag{S89}$$

As a concrete example, we take

$$\begin{aligned}
d_x^3(\boldsymbol{k}) &= T_{xy} \cos(k_x) \sin(k_y) + T_z \cos(2k_z), \\
d_y^3(\boldsymbol{k}) &= T_{xz} \sin(2k_x) \sin(k_z) + T_z \sin(2k_z), \\
d_z^3(\boldsymbol{k}) &= M + T_x \cos(2k_x) + T_y \cos(k_y).
\end{aligned} \tag{S90}$$

Here, $U_{\mathcal{G}_x}$ and $U_{\mathcal{G}_y}$ are both identity matrices. In main text Fig.1 (c), we plot a constant energy surface of this model, where parameters are taken as $E = -1.3, M = 0.5, T_x = 0.5, T_y = T_z = T_{xy} = T_{xz} = 1$.

### Second amphidicosm $\mathcal{M}^4$

The Hamiltonian with $k$-space $Pna2_1$ symmetry must satisfy

$$\begin{aligned}
H^4(k_x, k_y, k_z) &= U_{\mathcal{G}_x} H^3(-k_x, k_y + \pi, k_z + \pi) U_{\mathcal{G}_x}^\dagger, \\
H^4(k_x, k_y, k_z) &= U_{\mathcal{G}_y} H^3(k_x + \pi, -k_y, k_z) U_{\mathcal{G}_y}^\dagger,
\end{aligned} \tag{S91}$$

As a concrete example, we take

$$\begin{aligned} d_x^4(\bm{k}) &= T_{xy}\sin(k_x)\sin(k_y) + T_{yz}\cos(k_y)\cos(k_z), \\ d_y^4(\bm{k}) &= T_{xy}\cos(k_x)\sin(k_y) + T_{xz}\cos(2k_x)\sin(k_z), \\ d_z^4(\bm{k}) &= M + T_y\cos(2k_y) + T_z\cos(2k_z). \end{aligned} \qquad (S92)$$

Here, $U_{\mathcal{G}_x} = \sigma_1$ and $U_{\mathcal{G}_y}$ is an identity matrix. In main text Fig.1 (d), we plot a constant energy surface of this model, where parameters are taken as $E = -1.3, M = 0.5, T_y = T_z = 0.5, T_{xy} = T_{xz} = T_{yz} = 1$.

### Dicosm $\mathcal{M}^5$

The Hamiltonian with $k$-space $P2_1$ symmetry must satisfy

$$H^5(k_x, k_y, k_z) = U_{\mathcal{R}_2} H^5(-k_x, -k_y, k_z + \pi) U_{\mathcal{R}_2}^\dagger. \qquad (S93)$$

As a concrete example, we take

$$\begin{aligned} d_x^5(\bm{k}) &= T\sin(k_x + k_y)\cos(k_z) + T_x\cos(k_x) + T_y\cos(k_y), \\ d_y^5(\bm{k}) &= T_x\sin(k_x) + T_y\sin(k_y), \\ d_z^5(\bm{k}) &= M + T_z\cos(2k_z). \end{aligned} \qquad (S94)$$

Here, $U_{\mathcal{R}_2} = \sigma_1$. In main text Fig. 1(e), we plot a constant energy surface of this model, where parameters are taken as $E = -2.1, M = 0.5, T_x = 2, T_y = 1, T = T_z = 0.5$.

### Tricosm $\mathcal{M}^6$

The Hamiltonian with $k$-space $P3_1$ symmetry must satisfy

$$H^6(k_x, k_y, k_z) = U_{\mathcal{R}_3} H^6(-\frac{1}{2}k_x + \frac{\sqrt{3}}{2}k_y, -\frac{\sqrt{3}}{2}k_x - \frac{1}{2}k_y, k_z + 2\pi/3) U_{\mathcal{R}_3}^\dagger. \qquad (S95)$$

As a concrete example, we take

$$\begin{aligned} d_x^6(\bm{k}) &= \cos(\frac{2k_x}{\sqrt{3}})\cos(k_z) + \cos(\frac{k_x}{\sqrt{3}} - k_y)\cos(k_z + \frac{2\pi}{3}) + \cos(\frac{k_x}{\sqrt{3}} + k_y)\cos(k_z + \frac{4\pi}{3}), \\ d_y^6(\bm{k}) &= \sin(\frac{2}{\sqrt{3}}k_x)\cos(k_z) + \sin(\frac{k_x}{\sqrt{3}} - k_y)\cos(k_z + \frac{2\pi}{3}) + \sin(\frac{k_x}{\sqrt{3}} + k_y)\cos(k_z + \frac{4\pi}{3}), \\ d_z^6(\bm{k}) &= M + T_z\cos(3k_z). \end{aligned} \qquad (S96)$$

Here, $U_{\mathcal{R}_3}$ is an identity matrix. In main text Fig. 1(f), we plot a constant energy surface of this model, where parameters are taken as $E = -1.3, M = 0.5, T_z = 0.25$.

### Tetracosm $\mathcal{M}^7$

The Hamiltonian with $k$-space $P4_1$ symmetry must satisfy

$$H^7(k_x, k_y, k_z) = U_{\mathcal{R}_4} H^7(k_y, -k_y x k_z + \pi/2) U_{\mathcal{R}_4}^\dagger. \qquad (S97)$$

As a concrete example, we take

$$\begin{aligned} d_x^7(\bm{k}) &= T\sin(k_y)\sin(k_z) - T\sin(k_x)\cos(k_z), \\ d_y^7(\bm{k}) &= T\sin(k_y)\cos(k_z) - T\sin(k_x)\sin(k_z), \\ d_z^7(\bm{k}) &= M + T_z\cos(4k_z). \end{aligned} \qquad (S98)$$

Here, $U_{\mathcal{R}_4} = \sigma_1$. In main text Fig. 1(g), we plot a constant energy surface of this model, where parameters are taken as $E = -1, M = 0.5, T = 1, T_Z = 0.25$.

### Hexacosm $\mathcal{M}^8$

The Hamiltonian with $k$-space $P6_1$ symmetry must satisfy

$$H^8(k_x, k_y, k_z) = U_{\mathcal{R}_6} H^8(\frac{1}{2}k_x + \frac{\sqrt{3}}{2}k_y, -\frac{\sqrt{3}}{2}k_x + \frac{1}{2}k_y, k_z + \pi/3) U_{\mathcal{R}_6}^\dagger. \tag{S99}$$

As a concrete example, we take

$$\begin{aligned}
d_x^8(\boldsymbol{k}) &= \sin(\frac{2k_x}{\sqrt{3}})\cos(k_z) + \sin(\frac{k_x}{\sqrt{3}} + k_y)\cos(k_z + \frac{\pi}{3}) - \sin(\frac{k_x}{\sqrt{3}} - k_y)\cos(k_z + \frac{2\pi}{3}), \\
d_y^8(\boldsymbol{k}) &= \cos(\frac{2k_x}{\sqrt{3}}) + \cos(\frac{k_x}{\sqrt{3}} + k_y) + \cos(\frac{k_x}{\sqrt{3}} - k_y), \\
d_z^8(\boldsymbol{k}) &= M.
\end{aligned} \tag{S100}$$

Here, $U_{\mathcal{R}_6}$ is an identity matrix. In main text Fig. 1(h), we plot a constant energy surface of this model, where parameters are taken as $E = -1.3, M = 0.5$.

### Diciosm $\mathcal{M}^9$

The Hamiltonian with $k$-space $P2_12_12_1$ symmetry must satisfy

$$\begin{aligned}
H^9(k_x, k_y, k_z) &= U_{\mathcal{R}_z} H^9(-k_x, -k_y + \pi, k_z + \pi) U_{\mathcal{R}_z}^\dagger, \\
H^9(k_x, k_y, k_z) &= U_{\mathcal{R}_y} H^9(-k_x + \pi, k_y + \pi, -k_z) U_{\mathcal{R}_y}^\dagger.
\end{aligned} \tag{S101}$$

As a concrete example, we take

$$\begin{aligned}
d_x^9(\boldsymbol{k}) &= T_{xy} \cos(k_x)\sin(k_y) + T_{yz}\cos(k_y)\sin(k_z), \\
d_y^9(\boldsymbol{k}) &= T_{xy} \cos(k_x)\cos(k_y) + T_{yz}\sin(k_y)\sin(k_z), \\
d_z^9(\boldsymbol{k}) &= M + T_z \cos(2k_x).
\end{aligned} \tag{S102}$$

Here, $U_{\mathcal{R}_z} = \sigma_1$ and $U_{\mathcal{R}_y}$ is an identity matrix. In Fig. 1(i), we plot a constant energy surface of this model, where parameters are taken as $E = -1.3, M = 0.5, T_{xy} = T_{yz} = 1, T_z = 0.5$.

## VI. The tight-binding models

Besides the two-band models introduced above, we can also realize Brillouin platycosms in lattice models with proper flux configuration. We present possible lattice realization for all Biebarbach groups, except the trivial $P1$, in Fig. S10.

As illustrated in Fig. S10(a), (b), and (e), which correspond to $k$-space $Pc$, $Cc$, and $P2_1$, these models are constructed from monoclinic lattices, each containing 8 sites in the unit cell. As shown in Fig. S10(a), $\pi$ fluxes are distributed diagonally in the $x$-$z$ plane, while no fluxes are present in the other two directions. This configuration causes the mirror symmetry $M_x$ to anti-commute with the translation symmetry $L_z$. In Fig. S10(b), $\pi$ fluxes are arranged in a stripe distribution that passes through the $y$-$z$ plane. Similar to Fig. S10(a), this configuration also contributes to the anti-commutation of mirror symmetry and the translation symmetry $L_z$. The difference is that this lattice is base-centered monoclinic. The lattice model of $k$-space $P2_1$, as shown in Fig. S10(e), exhibits stripe-distributed $\pi$ fluxes through the $x$-$z$ plane. Thus, the 2-fold rotation symmetry around the $z$-axis anti-commutes with $L_z$.

Figures S10(c), (d), and (i) correspond to the $k$-space $Pca2_1$, $Pna2_1$, and $P2_12_12_1$, respectively. These models are constructed from primitive orthorhombic lattices, each containing 8 sites in the unit cell. The lattices shown in Figs. S10(c) and (d) both exhibit real space $Pmm2$ symmetry. The lattice model for $k$-space $Pca2_1$, depicted in Fig. S10(c), features diagonally distributed $\pi$ fluxes in the $x$-$z$ plane, as well as stripe-distributed $\pi$ fluxes through the $x$-$y$ plane. In contrast, Fig. S10(d), corresponding to $k$-space $Pna2_1$, displays $\pi$ fluxes arranged in a stripe distribution that passes through both the $x$-$z$ and $x$-$y$ planes. Additionally, the lattice model for $k$-space $P2_12_12_1$, illustrated in Fig. S10(i), has real space $P222$ symmetry and exhibits stripe-distributed $\pi$ fluxes through the $x$-$z$, $x$-$y$, and $y$-$z$ planes.

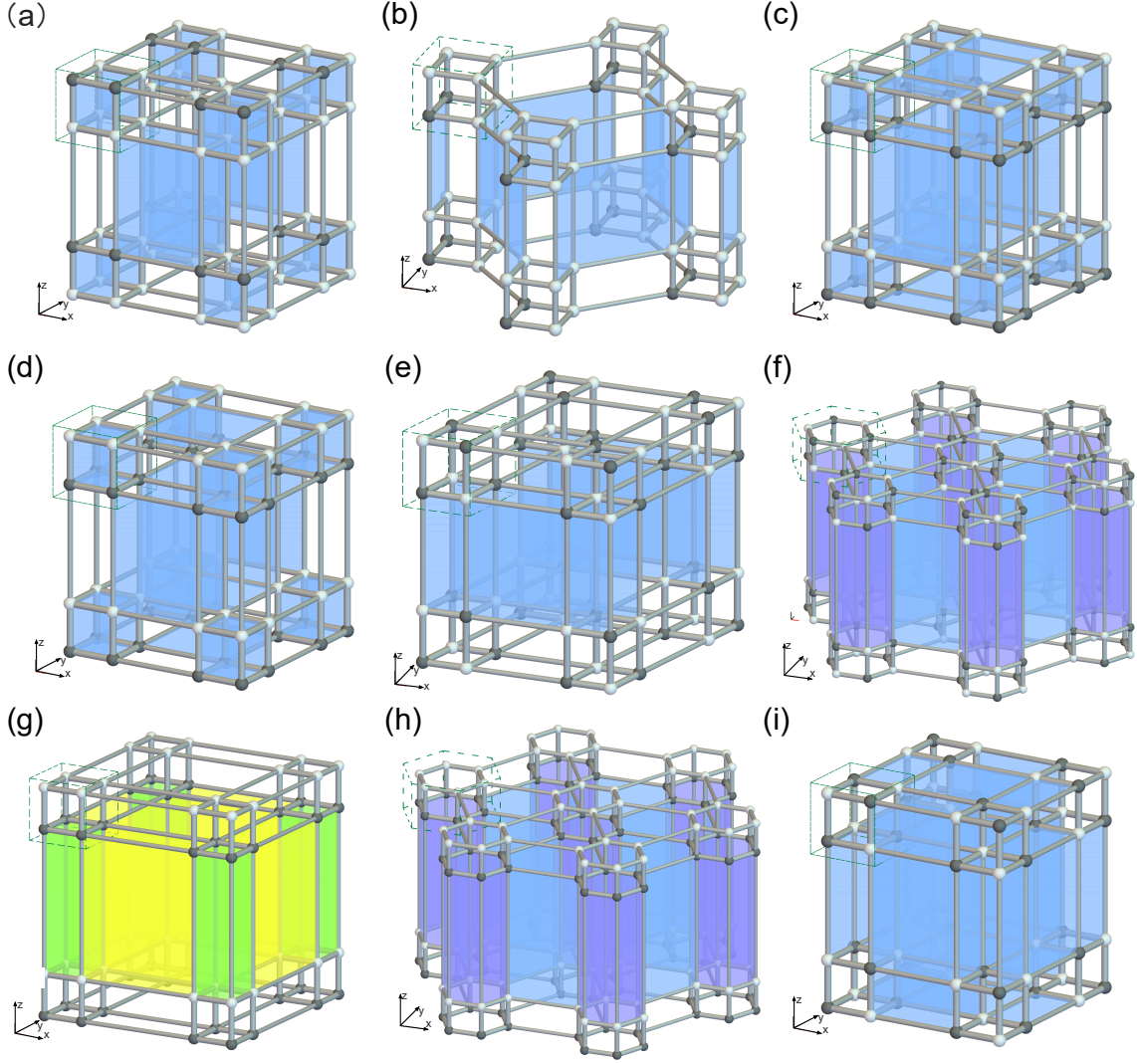

FIG. S10: The lattice models feature flux patterns that realize all $k$-space Biebarbach groups except $P1$. Specifically, these models correspond to (a) $k$-sapce $Pc$, (b) $k$-space $Cc$, (c) $k$-space $Pca2_1$, (d) $k$-space $Pna2_1$, (e) $k$-space $P2_12_12_1$, (f) $k$-space $P2_1$, (g) $k$-space $P3_1$, (h) $k$-space $P4_1$, (i) $k$-space $P6_1$. In each lattice model, the unit cell is outlined by a green dashed box. The blue-shaded regions in (a-f), (h) and (I) exhibit fluxes of $\pi$ mod $2\pi$. The purple shaded regions in (f) and (h) have fluxes $2\pi/3$. The yellow and green shaded regions in (g) represent fluxes $\pi/2$ and $3\pi/2$, respectively. Additionally, sites with different on-site energies are represented in two different colors.

The lattice models corresponding to $k$-space $P3_1$ and $P6_1$ are illustrated in Figs. S10(f) and (h). The model shown in Fig. S10(f) is constructed from a trigonal lattice and obeys real space $P3$ symmetry, while the model in Fig. S10(h) is based on a hexagonal lattice and adheres to real space $P6$ symmetry. Both models contain 12 sites in the unit cell and share the same flux configuration, with fluxes passing through planes parallel to the $z$-axis. The surfaces between cells in the $z$-direction, indicated in purple, contain a $2\pi/3$ flux. The regions labeled in blue represent $\pi$ fluxes. Therefore, 3-fold and 6-fold rotation symmetries now don't commute with translation symmetry $L_z$, instead they contribute to fractional translations in momentum space.

The last model of $k$-space $P4_1$ is illustrated in Fig. S10(g). This model is constructed from a tetragonal lattice and contains 8 sites in the unit cell. Fluxes are arranged in a stripe distribution across the $x$-$z$ and $y$-$z$ planes. As shown in Fig. S10(g), the regions labeled in green and yellow represent $\pi/2$ and $3\pi/2$ fluxes, respectively.

## VII. Weyl semimetals over Brillouin platycosms

From Tab. S1, we observe that the Brillouin platycosms $\mathcal{M}^\alpha$, where $\alpha = 1, 2, 3, 4$, are nonorientable, meaning there is no globally consistent orientation. Nonorientability allows all Weyl points in the fundamental domain to share the same relative chirality, and therefore the total chirality can be any even integer. To illustrate this, we take $k$-space $Pna2_1$ as an example.

The Bloch Hamiltonian for the lattice model shown in Fig. S10(d), which exhibits $k$-space $Pna2_1$ symmetry, can be written as

$$H(\boldsymbol{k}) = \begin{bmatrix} \lambda & [\chi_1^{x,+}]^* & 0 & \chi_1^{y,+} & \chi^{z,+} & 0 & 0 & 0 \\ \chi_1^{x,+} & \lambda & \chi_1^{y,-} & 0 & 0 & \chi^{z,-} & 0 & 0 \\ 0 & [\chi_1^{y,-}]^* & \lambda & \chi_1^{x,-} & 0 & 0 & \chi^{z,-} & 0 \\ [\chi_1^{y,+}]^* & 0 & [\chi_1^{x,-}]^* & \lambda & 0 & 0 & 0 & \chi^{z,+} \\ [\chi^{z,+}]^* & 0 & 0 & 0 & -\lambda & [\chi_2^{x,-}]^* & 0 & \chi_2^{y,+} \\ 0 & [\chi^{z,-}]^* & 0 & 0 & \chi_2^{x,-} & -\lambda & \chi_2^{y,-} & 0 \\ 0 & 0 & [\chi^{z,-}]^* & 0 & 0 & [\chi_2^{y,-}]^* & -\lambda & \chi_2^{x,+} \\ 0 & 0 & 0 & [\chi^{z,+}]^* & [\chi_2^{y,+}]^* & 0 & [\chi_2^{x,+}]^* & -\lambda \end{bmatrix}, \tag{S103}$$

where $\chi_a^{x,\pm} = \pm t_{a1}^x + t_{a2}^x e^{ik_x}$, $\chi_b^{y,\pm} = t_{b1}^y \pm t_{b2}^y e^{ik_y}$, and $\chi^{z,\pm} = t_1^z \pm t_2^z e^{ik_z}$, with $a, b = 1, 2$. To break time-reversal symmetry, we may include a perturbation term of the form $H^1(\boldsymbol{k}) = t\sigma_y \otimes \sigma_1 \otimes \sigma_1$. The bands touch at Weyl points when the parameters are chosen as follows: $t_{11}^x = 1.6$, $t_{12}^x = t_{11}^y = t_{12}^y = 1$, $t_{21}^x = t_{22}^y = t_1^z = 3$, $t_{22}^x = t_2^z = 1.5$, $M = 0$, and $t = 0.5$. Here we consider four valence bands.

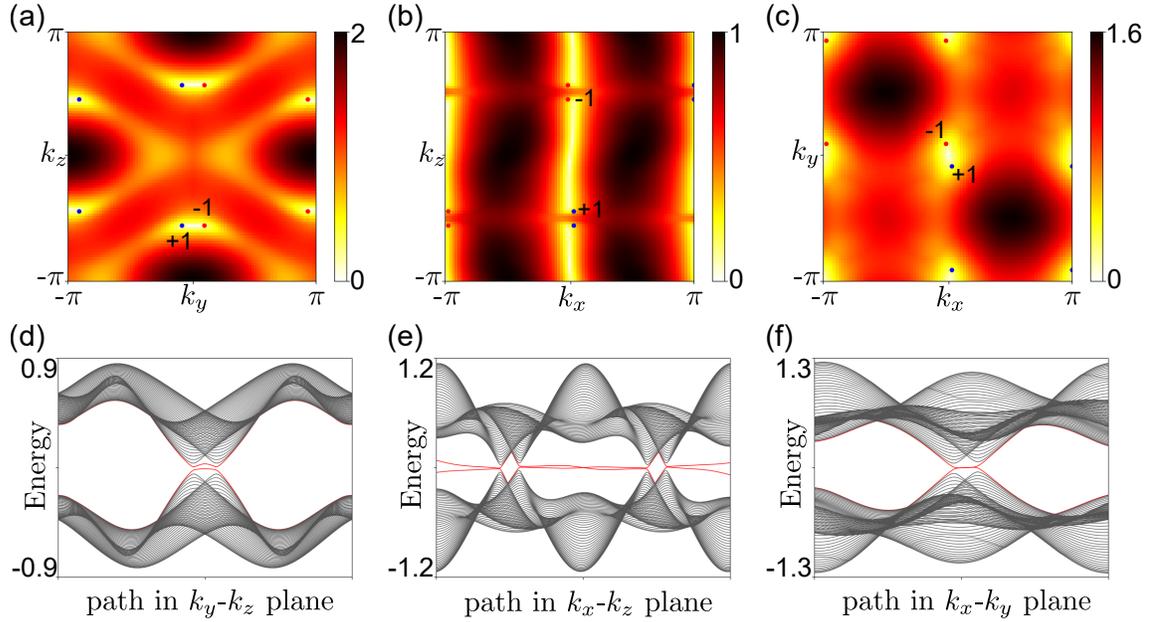

FIG. S11: Illustration of shapes and band structures of Fermi arcs. (a-c) show values of gaps between valence bands and conduction bands with open boundary conditions in $x, y$ and $z$ directions. Weyl points with chirality $\pm 1$ are labeled by blue(red) points respectively. In figures (d-f), corresponding band structures of Fermi arcs are presented.

The fundamental domain under the action of the $Pna2_1$ symmetry can be chosen as $[-\pi, \pi] \times [-\pi, 0] \times [-\pi, 0]$. As shown in Figs. S11(a), (b), and (c), there are 8 Weyl points in the entire Brillouin zone, with 2 Weyl points in the fundamental domain. The total chirality of the two Weyl points in the fundamental domain is $\chi = 2$.

When projected onto the surface Brillouin zone, Fermi arcs connect Weyl points with opposite chirality. For instance, when open boundary conditions(OBC) are applied in the $x$ direction, the Weyl points in the fundamental domain are projected into a region constrained by $-\pi \leq k_y \leq 0$ and $-\pi \leq k_z \leq 0$. The surface Hamiltonian exhibits momentum-space glide symmetry $\mathcal{G}_y$ and half-translation symmetry along the diagonal direction, denoted by $\mathcal{L}_{yz}$. Since the half-translation symmetry originates from the projection of glide symmetry $\mathcal{G}_x$, its action also changes the chirality of the Weyl points. As a result, the four Weyl points in the region $0 \leq k_y \leq \pi$, related to the Weyl points in the fundamental domain by $\mathcal{G}_y$ and $\mathcal{L}_{yz}$, have opposite chirality ($\chi = -1$). Also, similar arguments hold for cases with OBC in $y$ or $z$ directions, only with symmetry constraints different.

Fermi arcs connecting Weyl points of opposite chirality are also constrained by these symmetries, and their band structures are illustrated in Figs. S11(d), (e), and (f). Here, for each case, we present only one component of Fermi arcs, since different components are related by symmetries.


\end{document}